\documentclass[journal, 12pt, onecolumn, draftclsnofoot]{IEEEtran}
\usepackage{authblk}
\usepackage{url,cite}
\usepackage{amssymb,mathtools}
\usepackage{amsthm}
\usepackage{mathrsfs}
\usepackage{cuted}
\usepackage{tcolorbox}

\usepackage{graphicx}
\usepackage{multirow}

\usepackage{graphicx}         
\usepackage{booktabs}         

\usepackage{arydshln}
\usepackage{bm}
\usepackage{comment}
\includecomment{mycomment}

\newcommand\overmat[2]{%
  \makebox[0pt][l]{$\smash{\overbrace{\phantom{%
    \begin{matrix}#2\end{matrix}}}^{\text{$#1$}}}$}#2}

\usepackage{braket}

\usepackage{enumerate}

\usepackage{tikz}
\usetikzlibrary{tikzmark, shapes, fit,backgrounds}
\usetikzlibrary{matrix,decorations.pathreplacing,angles,quotes}
\usetikzlibrary{patterns, calc, positioning}
\usetikzlibrary{arrows,arrows.meta}

\tikzstyle{line} = [draw, -latex]

\usepackage{enumitem}
\setlist[enumerate]{wide = 0pt, leftmargin=*}
\usepackage{etoolbox} 
\makeatletter
\newif\if@gather@prefix 
\preto\place@tag@gather{%
  \if@gather@prefix\iftagsleft@ 
    \kern-\gdisplaywidth@ 
    \rlap{\gather@prefix}%
    \kern\gdisplaywidth@ 
  \fi\fi 
} 
\appto\place@tag@gather{%
  \if@gather@prefix\iftagsleft@\else 
    \kern-\displaywidth 
    \rlap{\gather@prefix}%
    \kern\displaywidth 
  \fi\fi 
  \global\@gather@prefixfalse 
} 
\preto\place@tag{%
  \if@gather@prefix\iftagsleft@ 
    \kern-\gdisplaywidth@ 
    \rlap{\gather@prefix}%
    \kern\displaywidth@ 
  \fi\fi 
} 
\appto\place@tag{%
  \if@gather@prefix\iftagsleft@\else 
    \kern-\displaywidth 
    \rlap{\gather@prefix}%
    \kern\displaywidth 
  \fi\fi 
  \global\@gather@prefixfalse 
} 
\def\math@cr@@@align{%
  \ifst@rred\nonumber\fi
  \if@eqnsw \global\tag@true \fi
  \global\advance\row@\@ne
  \add@amps\maxfields@
  \omit
  \kern-\alignsep@
  \if@gather@prefix\tag@true\fi
  \iftag@
    \setboxz@h{\@lign\strut@{\make@display@tag}}%
    \place@tag
  \fi
  \ifst@rred\else\global\@eqnswtrue\fi
  \global\lineht@\z@
  \cr
}
\newcommand*{\beforetext}[1]{%
  \ifmeasuring@\else
  \gdef\gather@prefix{#1}%
  \global\@gather@prefixtrue 
  \fi
} 
\makeatother

\newtheorem{lemma}{Lemma}
\newtheorem{theorem}{Theorem}

\newtheorem{proposition}{Proposition}
\newtheorem{remark}{Remark}

\theoremstyle{definition}
\newtheorem{definition}{Definition}
\newtheorem{protocol}{Protocol}

\DeclareMathOperator{\trace}{tr}

\DeclarePairedDelimiterX\symp[2]{\langle}{\rangle_\mathbb{S}}{#1 , #2}
\DeclarePairedDelimiterX\ipp[2]{\langle}{\rangle}{#1 , #2}

\newcommand{\bA}{\mathbf{A}}

\newcommand{\bG}{\mathbf{G}}
\newcommand{\bH}{\mathbf{H}}
\newcommand{\bI}{\mathbf{I}}

\newcommand{\bP}{\mathbf{P}}

\newcommand{\bZ}{\mathbf{Z}}
\newcommand{\ba}{\mathbf{a}}
\newcommand{\bb}{\mathbf{b}}
\newcommand{\bc}{\mathbf{c}}

\newcommand{\be}{\mathbf{e}}

\newcommand{\bs}{\mathbf{s}}

\newcommand{\bu}{\mathbf{u}}
\newcommand{\bv}{\mathbf{v}}
\newcommand{\bw}{\mathbf{w}}
\newcommand{\bx}{\mathbf{x}}

\newcommand{\bz}{\mathbf{z}}

\def\bzero{\mathbf{0}}

\def\F{\mathbb{F}}

\def\Fq{\F_{q}}

\begin{document}
\title{\Large Quantum $X$-Secure $T$-Private Information\\[0cm] Retrieval From MDS Coded Storage With\\[0cm] Unresponsive and Byzantine Servers}
\author{\normalsize Yuxiang Lu and Syed A. Jafar\\
{\small Center for Pervasive Communications and Computing (CPCC)}\\
{\small University of California Irvine, Irvine, CA 92697}\\
{\small \it Email: \{yuxiang.lu, syed\}@uci.edu}
\thanks{The results of this work were presented in part at IEEE ICC 2024 \cite{QEXSTPIR_ICC}. See Section \ref{sec:related} for details.}
}
\date{}
\maketitle

\begin{abstract}
A communication-efficient protocol is introduced over a  many-to-one quantum network for Q-E-B-MDS-X-TPIR, i.e., quantum private information retrieval  with MDS-$X$-secure storage and $T$-private queries. The protocol  is resilient to any set of up to $E$ unresponsive servers (erased servers or stragglers) and any set of up to $B$ Byzantine servers. The underlying coding scheme incorporates an enhanced version of a Cross Subspace Alignment (CSA) code, namely a Modified CSA (MCSA) code, into the framework of CSS codes. The error-correcting capabilities of CSS codes are leveraged to  encode the dimensions that carry desired computation results from the MCSA code into the error space of the CSS code, while the undesired interference terms are aligned into the stabilized code space. The challenge is to do this efficiently while also correcting quantum erasures and Byzantine errors. The protocol achieves superdense coding gain over comparable classical baselines for Q-E-B-MDS-X-TPIR, recovers as special cases the state of art results for various other quantum PIR settings previously studied in the literature, and paves the way  for applications in quantum coded distributed computation, where CSA code structures are important for communication efficiency, while security and resilience to stragglers and Byzantine servers are critical.
\end{abstract}

\begin{IEEEkeywords}
Coded storage, PIR, QMAC, security.
\end{IEEEkeywords}

\section{Introduction}
Recent interest in entanglement assisted computation over quantum many to one (also referred to as quantum multiple access (QMAC)) networks adds fundamentally novel dimensions to the  rapidly expanding theory of distributed communication and computation, beyond its classical cornerstones such as secret-sharing\cite{shamir1979share,cleve1999share,Q_StairCase_SS,MH_QSS}, private information retrieval (PIR)\cite{PIRfirstjournal,Sun_Jafar_TPIR,Freij_MDSTPIR,Tajeddine_Gnilke_Karpuk_Hollanti,QTPIR,QMDSTPIR,aytekin2023quantum}, coded distributed computation and computation networks \cite{Yu_Lagrange,Song_Hayashi_Secure_Q_Network_Coding,Yao_Jafar_QLCMAC}, and secure multiparty computation \cite{Damagard_MPC,PSQM,Christensen_Popovski_Quantum_Product,Lu_Yao_Jafar_KProd,Aytekin_Nomeir_Ulukus_QSecAgg}. Ideas from  these diverse perspectives are  encapsulated in a variety of specialized coding structures --- Reed-Solomon (RS) codes\cite{Macwilliams}, Cross Subspace Alignment (CSA) codes \cite{Jia_Jafar_MDSXSTPIR}, Lagrange Coded Computing \cite{Yu_Lagrange}, and CSS codes\cite{CSS_CS,CSS_S}, to name a few. Assimilating the specialized coding structures  is essential for a \emph{unified} theory that can facilitate a broader array of applications. This work represents such an endeavor, with the goal of developing a communication-efficient coding scheme (i.e., an efficient protocol) for  Q-E-B-MDS-X-TPIR \cite{Jia_Jafar_MDSXSTPIR}, i.e., quantum $X$-secure\footnote{$X$-security is  a secret-sharing constraint. The messages are the secret and the storage at each server is its share of the secret, such that any set of up to $X$ shares reveal nothing about the secret. There is another form of security, server secrecy \cite{QTPIR,QMDSTPIR,MH_QSS}, which requires that the user must not learn anything about any other message besides its desired message (also refered to as DB-privacy or symmetric privacy).  Note that $X$-security is not related to server secrecy, and that we consider only the former ($X$-security) in this work.} $T$-private information retrieval from MDS coded storage that is  resilient to up to $E$ unresponsive servers (equivalently referred to as erased servers) and up to $B$ Byzantine servers.\footnote{When assembled with `PIR', the abbreviation `Q' stands for `Quantum' (without `Q', the setting is classical by default), `E' stands for upto $E$ erased servers ( unresponsive servers), `B' stands for upto $B$ Byzantine servers, `MDS' stands for MDS coded storage, `X' stands for $X$-secure storage (so that up to $X$ colluding servers can learn nothing about the realizations of the {\bf stored messages}) and `T' stands for $T$-privacy constraint (so that up to $T$ colluding servers can learn nothing about which message is desired).}

\begin{figure}
\centering
\includegraphics{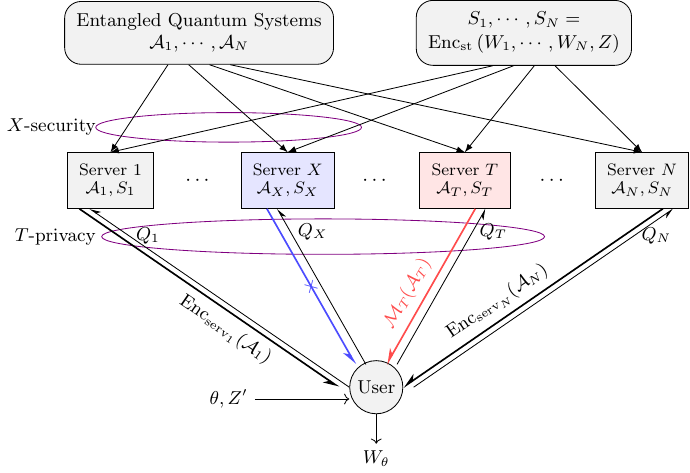}

\caption{Q-E-B-MDS-X-TPIR. Quantum systems $\mathcal{A}_1, \cdots, \mathcal{A}_{N}$ are prepared in an optimally entangled state and distributed to servers in advance. Messages $W_1, \cdots, W_K$, together with randomness $Z$ are encoded into $S_1, \cdots, S_N$ in an $X$-secure fashion and distributed to $N$ servers as their storage.  To privately retrieve a desired message $W_{\theta}$, $\theta\in[K]$, a user sends to the servers random (based on its its local randomness $Z'$) queries $Q_1, \cdots, Q_N$ that are $T$-private. Each server locally encodes its response into its quantum system  and sends it back to the user. In the figure, unresponsive (blue) server's quantum system is not received, and the Byzantine (red) server applies an arbitrary quantum channel to its quantum system.}
\label{fig:QEBMDSXSTPIR}
\end{figure}

In the Q-E-B-MDS-X-TPIR \cite{Jia_Jafar_MDSXSTPIR} setting as shown in Fig. \ref{fig:QEBMDSXSTPIR} there are $N$ servers equipped \emph{beforehand} (independent of the classical data) with optimally entangled quantum systems. Upon the  commencement of the protocol, there are $K$ \emph{classical} messages $W_1, \cdots, W_K$ (files, datasets) that are distributed among the servers in an MDS coded and $X$-secure fashion. MDS coding implies that the messages together with some classical randomness $Z$ (needed for security) are coded such that the storage size at each server is only a fraction $1/K_c$ of the original size of the $K$ messages. $X$-security means that  even if any set of up to $X$ servers collude they can learn nothing about the messages. A user (with its own private randomness $Z'$) wishes to efficiently retrieve the $\theta^{th}$ message ($\theta\in [K]$) by querying the $N$ servers in a $T$-private fashion. $T$-privacy means that even if any set of up to $T$ servers collude they can learn nothing about which message is desired by the user. The efficiency of the protocol is measured by the \emph{rate}, defined as the number of desired message bits retrieved per \emph{qubit} (a $d$-dimensional quantum system (sometimes called a \emph{qudit}) corresponds to $\log_2(d)$ qubits) of total download from the servers. Each server generates its response based on the user's queries and the storage available to that server, and encodes it into its own quantum system through local quantum operations. The quantum systems are then sent as answers from the servers to the user. The protocol must tolerate up to $E$ unresponsive servers, i.e., any set of up to $E$ servers may be unresponsive, equivalently their answers are erased over the QMAC. The protocol must also tolerate any set of up to $B$ Byzantine servers whose answers are subject to \emph{arbitrary} errors. Note that while the user's queries are sent without knowledge of which servers may turn out to be unresponsive, once the user receives the quantum systems in response, it knows which servers' answers were erased (known-position error), i.e, which servers did not respond. The identities of the Byzantine servers are not  directly revealed to the user from the answers. This corresponds to unknown-position errors in the context of error correcting codes. Resilience to unresponsive and Byzantine servers means that we require that regardless of which $E$ servers are erased, and which $B$ servers are Byzantine, the  protocol must allow the user to recover its desired message by measuring the received quantum systems.

Our solution centers around CSS codes and the classical CSA coding scheme  originally introduced for X-TPIR, i.e., PIR with $X$-secure storage and $T$-private queries \cite{Jia_Sun_Jafar_XSTPIR}, and subsequently applied to a number of classical variants of PIR, coded computing and private read-write designs for federated submodel learning \cite{PIR_tutorial}. The classical CSA scheme was  generalized to a quantum CSA scheme for Q-MDS-X-TPIR over the quantum many-to-one network in \cite{Allaix_N_sum_box,Lu_QCSA}, and its resilience to eavesdroppers was explored in \cite{aytekin2023quantum}.

\subsection{Challenges and Contributions}
While we focus on Q-E-B-MDS-X-TPIR to motivate the protocol developed in this work, we expect the protocol to be much more broadly relevant. This is because the underlying challenge is how to efficiently transmit CSA coded classical symbols when there are quantum resources shared among servers, some of which can be unresponsive (stragglers) and/or Byzantine. CSA code structures are not limited to PIR. For example, CSA codes feature prominently in the broad area of  \emph{coded distributed computation} (CDC) \cite{XSPPC,upload_vs_download_SDMM,Yu_Lagrange,Jia_Jafar_CDBC}. Thus,  the  protocol  from this work could potentially be a useful stepping stone towards future studies of quantum CDC (QCDC).\footnote{The  MDS storage can be viewed as coded matrix $A$, and the $T$ private queries as coded matrix $B$. The computation of $AB$ is distributed among servers. The MDS constraint limits upload cost, $X$-security/$T$-privacy protect against curious servers, and resilience to unresponsive/Byzantine servers guarantees robustness of the distributed computation.} Byzantine servers  are more challenging in the quantum setting, because the same quantum entanglement that allows gains in communication efficiency under ideal conditions, also makes entangled protocols more susceptible to stragglers and Byzantine adversaries, as their actions  impact not only their own quantum systems, but also the overall state of all entangled quantum systems. The challenges are listed as follows.

\begin{enumerate}[start=1,wide = 3pt, leftmargin = 0em]
	\item Compared with \cite{Allaix_N_sum_box,Lu_QCSA} that studied  Q-MDS-X-TPIR, the main challenge is to achieve resilience to unresponsive and Byzantine servers. In classical settings, this is done by having the answers form an error correcting code (ECC) of the desired message symbols (and interfering symbols introduced due to various constraints such as $X$-security, $T$-privacy and MDS storage) so that erasures or errors can be corrected first, after which the desired message symbols can be recovered. This idea is not directly applicable to quantum PIR schemes. Even though QPIR schemes are typically based on the stabilizer formalism \cite{QTPIR,QMDSTPIR,Allaix_N_sum_box}, the error-correcting capabilities of stabilizer codes are \emph{not} utilized to correct errors. Specifically, instead of the code space of a stabilizer code, in QPIR the information is encoded into the \emph{error space} \cite{Quantum_Error_Correction}, and is extracted by the user by measuring the qudits (quantum digits, a specific representation of quantum systems that will be explained in Section \ref{subsec:basic_pre}) with stabilizers to reveal the syndromes. Thus, the received $N$ answer qudits in QPIR are not in the stabilizer code space, even in the absence of erasures or errors.
	\item Compared with \cite{MH_QSS} that explored Q-TPIR with general access structure that involves resilience to $E$ unresponsive servers as a special case, the main challenge is to come up with an \emph{efficient} scheme that satisfies $X$-\emph{security} and \emph{MDS storage} constraints. Unlike the random coding based scheme that appears in \cite{MH_QSS}, the CSA code structure is important to accommodate $X$-\emph{security} and \emph{MDS storage}. Note that even in the classical setting, CSA codes allow higher communication rates in PIR with these two constraints (e.g., the CSA code based scheme \cite{Jia_Sun_Jafar_XSTPIR,Jia_Jafar_MDSXSTPIR} can achieve higher  rates than those achieved without CSA codes in \cite{XSPIR,XSPPC}). 
	\item Utilizing CSA codes further prevents us from placing the answering qudits in the code space of a stabilizer code (without considering the erasure or Byzantine errors). Specifically, the CSA code is the direct sum of a Reed-Solomon code of \emph{interfering/undesired} symbols and a Cauchy RS code of \emph{desired} symbols. It is non-trivial to construct a CSS code upon two CSA codes $\mathrm{CSA}_X, \mathrm{CSA}_Z$, such that $\mathrm{CSA}_X^\perp \subset \mathrm{CSA}_Z$. This is because the dual code of a CSA code should be dual to both the RS part and the Cauchy RS part, whose structures are not trivially compatible. 
\end{enumerate}
Thus, our main \emph{contribution} is a protocol that utilizes the error-correcting capabilities of CSS codes, i.e., the information carrying ability of their syndromes as the underlying framework. Within this framework, the protocol exploits the RS sub-code of CSA codes to efficiently retrieve the desired computation results (desired message symbols in the PIR problem) that are encoded by \emph{classical} codes,\footnote{We refer to the desired message symbols as the computation results to emphasize that they are the outcome of the \emph{computation task}, e.g., PIR. } while also tolerating \emph{quantum} erasures and Byzantine errors. Intuitively, in the underlying classical CSA code based protocol,  the answers from the servers are viewed as the RS sub-code of \emph{interfering} symbols, with Cauchy RS code of \emph{desired} message symbols added as ``error.'' The syndrome of the RS sub-code  uniquely identifies the ``error'' in the Cauchy RS code space together with the actual errors introduced by unresponsive or Byzantine servers. From the quantum perspective, the shared qudits are initially in the code space of the CSS code constructed from the RS sub-codes of two instances of CSA codes. Servers apply Pauli operators to their qudits to encode the answers generated according to the two instances of the CSA code based classical scheme. The Pauli operators' components that correspond to RS sub-codes of interfering symbols are not detectable since they commute with stabilizers, while the component corresponding to desired message symbols, together with the errors introduced by unresponsive and Byzantine servers, are identified through syndrome measurement. In a nutshell, dimensions that carry desired computation results from the CSA code are encoded into the error space of the CSS code, while the undesired interference terms are aligned into the stabilized code space. A technicality worth noting is that a key enhancement is made to the CSA code, transforming it into a Modified CSA (MCSA) code --- whereby the RS sub-code is turned into a GRS sub-code whose dual code is still a GRS code, so that a CSS code can be easily constructed on $\mathrm{GRS}_X, \mathrm{GRS}_Z$ that are sub-codes of two MCSA codes, where $\mathrm{GRS}_X^\perp \subset \mathrm{GRS}_Z$. This `MCSA-CSS' construction can be found in Protocol \ref{proto:QCSA-CSS} in this work.

While there is entanglement shared beforehand among the distributed servers (transmitters), it is important to note that the servers  do \emph{not} share any entanglement in advance with the user (the receiver). Intuitively, the shared entanglement among transmitters leads to a  superdense coding gain  in quantum PIR schemes allowing them to achieve in some cases twice the rate of their classical counterparts \cite{QTPIR,QMDSTPIR,Allaix_N_sum_box}. The quantum scheme proposed in this paper also achieves the factor of $2$ superdense coding gain  compared with the classical scheme proposed in \cite{Jia_Jafar_MDSXSTPIR}. It is also noteworthy that the quantum PIR setting addressed in this paper recovers as \emph{special cases} various other settings considered in the literature,  such as Q-B-X-TPIR in \cite{Nomeir_Aytekin_Ulukus_QBXSTPIR}, Q-E-TPIR in \cite{MH_QSS}, Q-MDS-X-TPIR in \cite{Allaix_N_sum_box}, Q-MDS-TPIR in \cite{QMDSTPIR}, and Q-TPIR in \cite{QTPIR}. Indeed, the protocol presented in this work achieves the state-of-the-art rates across all of the aforementioned special case scenarios.

\subsection{Comparison to related works}\label{sec:related}
The most closely related work is the conference version of this paper in \cite{QEXSTPIR_ICC,QEXSTPIR_arXiv} where Q-E-X-TPIR problem is studied  based on the $N$-sum box abstraction of \cite{Allaix_N_sum_box}. The conference version  allows neither MDS storage nor resilience to Byzantine servers. The conference version was then developed into a preliminary ArXiv version \cite{Lu_Jafar_QEBXSTPIR} of this paper where the approach taken for resilience to Byzantine servers that apply arbitrary Pauli errors is to guess the identities of Byzantine servers, treat them as erasures and decode, and check if there exists a set of decoding results that agree. However,  the resilience to \emph{arbitrary} Byzantine errors (rather than just Pauli errors)  is not explicit under the $N$-sum box abstraction. The present version further develops our approach, making the Byzantine resilience explicit. Instead of the $N$-sum box abstraction, here we directly utilize the fact that the syndrome measurement of a CSS code can reduce \emph{arbitrary} errors (that affect fewer qudits than its minimum distance) to Pauli errors (Lemma \ref{lem:error_reduction}).

Let us also note the parallel and independent work in \cite{Nomeir_Aytekin_Ulukus_QBXSTPIR} that  studies Q-B-X-TPIR  through the lens of the $N$-sum box abstraction, as further evidence of interest in this problem.

\subsection{Organization}
Section \ref{sec:preliminaries} introduces the notation together with some basic concepts of quantum systems, classical error correcting codes and quantum information. Section \ref{sec:problem_statement} formalizes the Q-E-B-MDS-X-TPIR problem. Section \ref{sec:main} presents our main result as Theorem \ref{thm:QEB}. Section \ref{sec:csa} revisits the CSA code based classical E-B-MDS-X-TPIR scheme which is crucial to  our construction. A modified CSA code (MCSA code) is presented in Section \ref{sec:QCSA}. The quantum protocol, namely MCSA-CSS, that builds upon the MCSA code and a CSS code, is presented in Section \ref{sec:QEB}. Section \ref{sec:conclusion} concludes the paper.

\section{Preliminaries}\label{sec:preliminaries}
\subsection{Miscellaneous}\label{subsec:basic_pre}
For two integers $a,b$, the set $\{a,a+1, \cdots, b\}$ is denoted as $[a:b]$. For compact notation, $[1:b]$ is denoted as $[b]$. For a set $\mathcal{S}$, $|\mathcal{S}|$ denotes its cardinality, and for any $k \leq |\mathcal{S}|, \binom{\mathcal{S}}{k} \triangleq \{S \mid  S \subset \mathcal{S}, |S| = k\}$. For an $r\times c$ matrix $\bA$, $\bA(\mathcal{A}, \mathcal{B})$ denotes the sub-matrix of $\bA$ whose row indices are in $\mathcal{A}$ and column indices in $\mathcal{B}$. $\mathcal{A}$ or $\mathcal{B}$ will be replaced by `$:$' if they contain all the  rows or columns,  respectively. If $\bA$ is a vector, we simply write $\bA(\mathcal{S})$ to denote the sub-vector of $\bA$ whose indices are in $\mathcal{S}$.
For two column vectors $\bc_1, \bc_2$, $[\bc_1;\bc_2] \triangleq [\bc_1^\top~~\bc_2^\top]^\top$, i.e., a longer column vector with $\bc_1$ stacked above $\bc_2$. $\mbox{colspan}(\bA)$ denotes the vector subspace spanned by the columns of $\bA$. If $\bA$ is a projection matrix, then $\mathrm{Im}(A) = \mbox{colspan}(\bA)$. $\mbox{ker}(\bA)$ is the kernel space of $\bA$. $\bA^{\dagger}$ is the conjugate transpose of $\bA$. For a length $n$ vector $\bv = [v_1~~v_2~~\cdots~~v_n]^\top$, $\mbox{Diag}(\bv)$ denotes the diagonal $n\times n$ matrix whose diagonal elements are entries of $\bv$. $\mathrm{supp}(\bv) \triangleq \{i \mid v_i \neq 0\}$ and $\mathrm{wt}(\bv) \triangleq |\mathrm{supp}(\bv)|$. $\bI_N$ is the $N\times N$ identity matrix. For any random variable  that is written in upper case (say, $Z$), we use the corresponding lower case ($z$) to denote its realization. The state of a quantum system $A$ defined on Hilbert space $\mathcal{H}_A$ is represented by a density operator $\rho_A \in \mathcal{D}_A$ where $\mathcal{D}_A$ is a set of all positive semi-definite operators with trace $1$ acting on $\mathcal{H}_A$. A pure state can also be represented by a unit vector in $\mathcal{H}_{A}$. For a classical-quantum system $XA$, $\rho_{A\mid X=x}$, or simply $\rho_{A\mid x}$, denotes the density operator of $A$ conditioned on the realization $X = x$. The label of the quantum system in the subscript may be omitted for compact notation if it is clear from the context. $\Fq$ is a finite field with order $q$ where $q = p^r$ is a prime power. The field trace $\trace_{\Fq/\mathbb{F}_p}(\cdot) \colon \Fq \rightarrow \mathbb{F}_p$  is an $\mathbb{F}_p$-linear map from $\Fq$ to $\mathbb{F}_p$, and  $\omega \triangleq e^{2\pi\sqrt{-1}/p}$. If a quantum system $A$ has dimension $|A| = q$, with $\{\ket{a}\}_{a \in \Fq}$ being its computational basis, we call it a $q$-dimensional qudit. 

\subsection{Classical Error Correcting Codes}
\begin{definition}
	{$[n,k,d]$ Code}: An $[n,k,d]$ classical code over $\Fq$ is the \textbf{column space} of a rank $k$ generator matrix $\bG \in \Fq^{n \times k}$, i.e., $\mathcal{C} = \mathrm{colspan}(\bG)$. It has a rank $n-k$ parity-check matrix $\bH \in \Fq^{n\times n-k}$ such that $\bH^\top\bG = \bzero$. The dual code of $\mathcal{C}$ is $\mathcal{C}^\perp = \mathrm{colspan}(\bH)$. If an $[n,k,d]$ code satisfies $d = n-k+1$, we  call it an $[n,k]$ MDS (maximum distance separable) code.
\end{definition}

\begin{definition}\label{def:GRS}
	{GRS Code}: A Generalized Reed-Solomon Code $\mathcal{C} = \mathrm{GRS}_{n,k}^{q,(\bm{\alpha}, \bu)}$ over $\Fq$ is the column space of the generator matrix defined in \eqref{eq:GRS_G} where $\bm{\alpha} = (\alpha_1, \alpha_2, \cdots, \alpha_n)$ are $n$ distinct elements in $\Fq$ and $\bu = (u_1, u_2, \cdots, u_n)$ are $n$ non-zero elements in $\Fq$. By definition, $q \geq n$.
	\begin{align}
		\bG_{\mathrm{GRS}_{n,k}^{q,(\bm{\alpha}, \bu)}} \triangleq
		\left[
			\begin{array}{cccc}
				u_1    & u_1 \alpha_{1} & \cdots & u_1 \alpha_{1}^{k-1} \\
				u_2    & u_2 \alpha_{2} & \cdots & u_2 \alpha_{2}^{k-1} \\
				\vdots & \vdots         & \vdots & \vdots               \\
				u_n    & u_n \alpha_{n} & \cdots & u_n \alpha_{n}^{k-1}
			\end{array}\right]\label{eq:GRS_G}
	\end{align}
\end{definition}

\begin{definition}\label{def:CRS}
	{CRS Code}: A Cauchy Reed-Solomon Code $\mathcal{C} = \mathrm{CRS}_{n,k}^{q,(\bm{\alpha}, \mathbf{f}, \bu)}$ over $\Fq$ is the column space of the generator matrix defined in \eqref{eq:CRS_G} where $\left(\bm{\alpha}, \mathbf{f}\right) = (\alpha_1, \alpha_2, \cdots, \alpha_n, f_1, f_2, \cdots, f_k)$ are $n+k$ distinct elements and $\bu = (u_1, u_2, \cdots, u_n)$ are $n$ non-zero elements. By definition, $q \geq n+k$.
	\begin{align}
		\bG_{\mathrm{CRS}_{n,k}^{q,(\bm{\alpha}, \mathbf{f}, \bu)}} \triangleq
		\left[
			\begin{array}{cccc}
				\frac{u_1}{f_1 - \alpha_1} & \frac{u_1}{f_1 - \alpha_1} & \cdots & \frac{u_1}{f_{k} - \alpha_1} \\
				\frac{u_2}{f_1 - \alpha_2} & \frac{u_2}{f_1 - \alpha_2} & \cdots & \frac{u_2}{f_{k} - \alpha_2} \\
				\vdots                     & \vdots                     & \vdots & \vdots                       \\
				\frac{u_n}{f_1 - \alpha_n} & \frac{u_n}{f_1 - \alpha_n} & \cdots & \frac{u_n}{f_{k} - \alpha_n}
			\end{array}\right]\label{eq:CRS_G}
	\end{align}
\end{definition}

\subsection{Quantum Information}
\begin{definition}
	{Quantum Channel}: A quantum channel with input quantum system $A$ and output quantum system $B$ is a completely positive trace preserving mapping (CPTP) $\mathcal{M} \colon \mathcal{D}_A \rightarrow \mathcal{D}_B$. It can be represented by Kraus Operators $\{K_i\}$ such that $\sum_i K_i^{\dagger}K_i$ is an identiy matrix and $\mathcal{M}(\rho) = \sum_{i} K_i \rho K_i^{\dagger}$.
\end{definition}
\begin{definition}
	{Pauli Operators for Qudits}\cite{Ketkar06}: For any $a, b \in \Fq$, define the single qudit Pauli Operators $\mathsf{X}^{b}, \mathsf{Z}^{b} \in \mathbb{C}^{q\times q}$ so that
	\begin{align}
		\mathsf{X}^{b}\ket{a} = \ket{a+b}, &  & \mathsf{Z}^{b}\ket{a} = \omega^{\trace_{\Fq/\mathbb{F}_p}(ba)}\ket{a}.\notag
	\end{align}
	For $n \in \mathbb{N}$ and any $\bx = [x_1~~\cdots~~x_n]^\top, \bz = [z_1~~\cdots~~z_n]^\top \in \Fq^{n\times 1}$, let the $n$-qudit Pauli Operators be defined as
	\begin{align}
		\mathsf{X}^{\bx}\mathsf{Z}^{\bz} \triangleq \bigotimes_{i \in [n]}\mathsf{X}^{x_i}\mathsf{Z}^{z_i}.\notag
	\end{align}
	Note that
	\begin{align}
		\left(\mathsf{X}^{\bx}\mathsf{Z}^{\bz}\right)\left(\mathsf{X}^{\bx'}\mathsf{Z}^{\bz'}\right) & = \omega^{\trace_{\Fq/\mathbb{F}_p}\left(\bz^\top\bx' - \bx^\top\bz'\right)}\left(\mathsf{X}^{\bx'}\mathsf{Z}^{\bz'}\right)\left(\mathsf{X}^{\bx}\mathsf{Z}^{\bz}\right)\notag \\
		                                                                                             & = \omega^{\trace_{\Fq/\mathbb{F}_p}\left(\bz^\top\bx'\right)}\mathsf{X}^{\bx+\bx'}\mathsf{Z}^{\bz+\bz'} \label{eq:XZ_mul}
	\end{align}
\end{definition}

\begin{definition}\label{def:CSS}
	{CSS Code\cite{CSS_CS,CSS_S,QLRC}}: A $\mathcal{C} = \mathrm{CSS}(\mathcal{C}_X,\mathcal{C}_Z)$ code encodes the state space of $k$ $q$-dimensional qudits  into a code space of $n$ $q$-dimensional qudits
	\begin{align}
		\mathrm{CSS}\left(\mathcal{C}_X,\mathcal{C}_Z\right) = \mathrm{colspan} \left(\sum_{\bx^\perp \in \mathcal{C}_X^\perp}\ket{\bx^\perp + \bz} \mid \bz \in \mathcal{C}_Z\right),
	\end{align}
	where $\mathcal{C}_X$, $\mathcal{C}_Z$ are classical $[n,k_X,d_X], [n,k_Z,d_Z]$ linear codes with generator matrices $\bG_{\mathcal{C}_X} \in \Fq^{n\times k_X}, \bG_{\mathcal{C}_Z} \in \Fq^{n \times k_Z}$ respectively, that satisfy $\mathcal{C}_X^\perp \subset \mathcal{C}_Z$. The $\mathrm{CSS}\left(\mathcal{C}_X,\mathcal{C}_Z\right)$ is a stabilizer code with stabilizers $S = \left\{\mathsf{X}^{\ba} \mathsf{Z}^{\bb} \mid \ba \in \mathcal{C}_X^\perp, \bb \in \mathcal{C}_Z^\perp\right\}$. Its minimum distance is $d \geq \min(d_X, d_Z)$.
\end{definition}

\begin{definition}\label{def:stabilizer_measure}
	{Stabilizer Measurement:} For the CSS code in Definition \ref{def:CSS}, for any $\ba \in \mathcal{C}_X^\top, \bb \in \mathcal{C}_Z^\top$, according to \cite[Appendix C, Fact 2)]{QTPIR}, the stabilizer $\mathsf{X}^{\ba} \mathsf{Z}^{\bb}$ can be decomposed as
	\begin{align}
		\mathsf{X}^{\ba} \mathsf{Z}^{\bb} = \sum_{i \in \mathbb{F}_p} \omega^i \bP_i^{\ba,\bb}
	\end{align}
	where $\left\{\bP_{i}^{\ba,\bb}\right\}_{i \in \mathbb{F}_p}$ are orthogonal projections such that
	\begin{align}
		\bP_{i}^{\ba,\bb}\bP_{j}^{\ba,\bb} = \bzero &&\forall i \neq j, \\
		\sum_{i \in \mathbb{F}_p} \bP_{i}^{\ba,\bb} = \bI.
	\end{align}
	Then the stabilizer measurement $\mathsf{X}^{\ba} \mathsf{Z}^{\bb}$ is defined as the \emph{Projection-Valued Measurement} (PVM, \cite{Nielson_Chuang}) with projections $\left\{\bP_{i}^{\ba,\bb}\right\}_{i \in \mathbb{F}_p}$.
\end{definition}

\begin{definition}\label{def:syndrome_measure}
	{Syndrome Measurement:} For the CSS code defined in Definition \ref{def:CSS}, a syndrome measurement is the stabilizer measurement corresponding to all the (generator) stabilizers according to Definition \ref{def:stabilizer_measure}.
\end{definition}

\begin{proposition}\label{prop:css_syndrome}
	(Well known) For any $\ket{\psi} \in \mathrm{CSS}\left(\mathcal{C}_X,\mathcal{C}_Z\right)$ and any $\bx, \bz \in \Fq^{n\times 1}$, the $n$-qudit pure state $\mathsf{X}^{\bx}\mathsf{Z}^{\bz}\ket{\psi}$ is an eigenvector for all the stabilizers, and its syndrome measurement outcome is as follows, with $\bH_{\mathcal{C}_X}$, $\bH_{\mathcal{C}_Z}$ being parity-check matrices for $\mathcal{C}_X, \mathcal{C}_Z$ respectively.
	\begin{align}
		\bs_X = \bH_{\mathcal{C}_Z}^\top \bx,~~\bs_Z = \bH_{\mathcal{C}_X}^\top \bz
	\end{align}
\end{proposition}

The following lemma will be useful.
\begin{lemma}\label{lem:error_reduction}
	Consider any $n$-qudit state $\ket{\psi} \in \mathrm{CSS}\left(\mathcal{C}_X,\mathcal{C}_Z\right)$ with the $n$ qudits labeled as $\mathcal{A}_{[n]}$. For any $\bx, \bz \in \Fq^{n\times 1}, \mathcal{S} \subset [n], |\mathcal{S}| \leq \min(d_X, d_Z)-1$ where $d_X, d_Z$ are distances of $\mathcal{C}_X, \mathcal{C}_Z$ respectively, suppose the $n$-qudit Pauli gate $\mathsf{X}^{\bx}\mathsf{Z}^{\bz}$ is first applied to $\mathcal{A}_{[n]}$. Then for any quantum channel $\mathcal{M}_{\mathcal{S}} \colon \mathcal{D}_{\mathcal{A}_{\mathcal{S}}} \rightarrow \mathcal{D}_{\mathcal{A}_{\mathcal{S}}}$ that is applied to qudits $\mathcal{A}_{\mathcal{S}}$, the syndrome measurement reduces the quantum channel to some Pauli operators only affecting qudits $\mathcal{A}_{\mathcal{S}}$, i.e., 
	\begin{align}
		 &\forall \ket{\psi} \in \mathrm{CSS}\left(\mathcal{C}_X,\mathcal{C}_Z\right); \bx, \bz \in \Fq^{n\times 1};\notag\\
		&\mathcal{S} \subset [n], |\mathcal{S}| \leq \min(d_X, d_Z)-1, \mathcal{M}_{\mathcal{S}} \colon \mathcal{D}_{\mathcal{A}_{\mathcal{S}}} \rightarrow \mathcal{D}_{\mathcal{A}_{\mathcal{S}}}, \notag\\
		 & \Big(\mathrm{id}_{[n]\setminus\mathcal{S}}\otimes\mathcal{M}_{\mathcal{S}}\Big)\left(\mathsf{X}^{\bx}\mathsf{Z}^{\bz}\ket{\psi}\bra{\psi}\left(\mathsf{X}^{\bx}\mathsf{Z}^{\bz}\right)^\dagger\right)\notag \\
		 & \overset{\mathrm{synd. meas.}}{\longrightarrow}
		 \mathsf{X}^{\bm{\epsilon}_\mathcal{S}^X}\mathsf{Z}^{\bm{\epsilon}_{\mathcal{S}}^Z}\left(\mathsf{X}^{\bx}\mathsf{Z}^{\bz}\ket{\psi}\bra{\psi}\left(\mathsf{X}^{\bx}\mathsf{Z}^{\bz}\right)^{\dagger}\right)\left(\mathsf{X}^{\bm{\epsilon}_\mathcal{S}^X}\mathsf{Z}^{\bm{\epsilon}_{\mathcal{S}}^Z}\right)^\dagger\notag\\
		 & \overset{\eqref{eq:XZ_mul}}{=}\mathsf{X}^{\bx + \bm{\epsilon}_\mathcal{S}^X}\mathsf{Z}^{\bz + \bm{\epsilon}_{\mathcal{S}}^Z}\ket{\psi}\bra{\psi}\left(\mathsf{X}^{\bx+\bm{\epsilon}_\mathcal{S}^X}\mathsf{Z}^{\bz+\bm{\epsilon}_\mathcal{S}^Z}\right)^{\dagger}, 
	\end{align}
	with the outcome being
	\begin{align}
		 & \bs_X = \bH_{\mathcal{C}_Z}^\top \left(\bx + \bm{\epsilon}_\mathcal{S}^X\right),~~\bs_Z = \bH_{\mathcal{C}_X}^\top \left(\bz + \bm{\epsilon}_\mathcal{S}^Z\right),
	\end{align}
	where $\mathrm{supp}(\bm{\epsilon}_{\mathcal{S}}^X) = \mathrm{supp}(\bm{\epsilon}_{\mathcal{S}}^Z) = \mathcal{S}$.
\end{lemma}
	Though the lemma is conceptually somewhat standard, we provide a proof in Appendix \ref{app:proof_error_reduction} for the sake of completeness.

\section{Problem Statement}\label{sec:problem_statement}
Let us start with the classical setting defined in \cite{Jia_Jafar_MDSXSTPIR}. There are $K$ messages $W_1,  \cdots, W_K$ that are i.i.d. uniform over $[\mathsf{M}]$. They are securely encoded with  randomness $Z\in\mathcal{Z}$ to form the storage at $N$ servers. For $\theta\in[K]$, the user wishes to privately retrieve the message $W_\theta$ by querying the $N$ servers. Local randomness $Z'\in\mathcal{Z}'$ is available to the user to generate private queries. For any $n \in [N]$, the \emph{random variables} regarding the storage, query and answer (in the classical setting) at server $n$, denoted as $S_n, Q_n^{[\theta]}$ and $A_n$ with realizations being $s_n, q_n, a_n$, are deterministic functions of the following $3$ independent random variables, whose realizations will be denoted as $w_{[K]},z,z'$ respectively.
\begin{align}
	 & [\mathrm{Messages}]: W_{[K]} \in [\mathsf{M}]^K,\notag  \\
	 & [\mathrm{Storage~Randomness}]: Z \in \mathcal{Z},\notag \\
	 & [\mathrm{User~Randomness}]: Z' \in \mathcal{Z}'.
\end{align}
The classical problem is similar to the quantum problem in Fig. \ref{fig:QEBMDSXSTPIR}, but there are no entangled quantum systems shared among servers and the answers from servers are classical symbols. Byzantine servers will return arbitrary classical symbols. Next we specify the storage, queries, servers' answers, and the user's decoding for both classical and quantum settings.

\subsection{Classical Setting}
\noindent\textbf{MDS and $X$-Secure Storage:} The storage at server $n, n\in [N]$ is denoted as $S_n \in [\mathsf{S}]$. With encoding function $\mathrm{Enc}_{\mathrm{st}}\colon [\mathsf{M}]^K \times \mathcal{Z}\rightarrow [\mathsf{S}]^N$, the storage $S_{[N]} = \mathrm{Enc}_{\mathrm{st}}(W_{[K]}, Z)$ forms an $[N, X + K_c]$ MDS code, such that
\begin{align}
	& \left[\mbox{MDS  Storage}\right] &  & H(W_{[K]} \mid S_\mathcal{S}) = 0,\notag                              \\
	&                                  &  & \hspace{1.5cm}\forall \mathcal{S} \subset [N], |\mathcal{S}| = X+K_c, \\
	&                                  &  & H(S_n) = \log_{2}\mathsf{S} = K\log_{2}(\mathsf{M})/K_c,\notag        \\
	&                                  &  & \hspace{1.5cm} \forall n \in [N]                                      \\
	& \left[X-\mbox{Security}\right]   &  & I(W_{[K]}; S_{\mathcal{X}}) = 0,\notag                                \\
	&                                  &  & \hspace{1.5cm} \forall \mathcal{X} \subset [N], |\mathcal{X}| \leq X.
\end{align}
i.e., any $X + K_c$ servers must be able to recover all the $K$-messages, the storage size at each server is $1/K_c$ of the total size of the $K$ messages, and any $X$ or fewer servers can learn nothing about the messages. The encoding is done by, e.g., sources of the messages.

\begin{remark}
	The storage  forms a ramp secret sharing \cite{Yamamoto_rampSS} of the $K$-message database. We call it MDS and secure storage for  comparison with Quantum MDS-PIR \cite{QMDSTPIR}, as when $X=0$, the above entropic constraints hold for an $[N,K_c]$ MDS code where $K_c$ message symbols are encoded into $N$ codeword symbols such that any $K_c$ codeword symbols recover the message and each codeword symbol is $1/K_c$ of the message size (since there are $K_c$ message symbols). When $K_c = 1$, there is no MDS storage constraint.
\end{remark}

\noindent\textbf{Queries:} A user wishes to retrieve the $\theta^{th}, \theta \in [K]$, message $W_{\theta}$ from the servers by sending the $T$-private queries $Q_1^{[\theta]}, Q_2^{[\theta]}, \cdots, Q_N^{[\theta]} \in \mathcal{Q}$ to the $N$ servers such that any $T$ or fewer servers learn nothing about $\theta$. Mathematically, using the encoding function $\mathrm{Enc}_{\mathrm{user}} \colon [K] \times \mathcal{Z}' \rightarrow \mathcal{Q}^N$, the user generates queries,
\begin{align}
	(Q_1^{[\theta]}, Q_2^{[\theta]}, \cdots, Q_N^{[\theta]}) = \mathrm{Enc}_{\mathrm{user}}\left(\theta, Z'\right)
\end{align}
where $Z' \in \mathcal{Z}'$ is the user's local randomness. Meanwhile, the $T$-privacy constraint must be satisfied such that
\begin{align}
	&\left[T-\mbox{Privacy}\right] &  & \left(S_\mathcal{T}, Q_{\mathcal{T}}^{[\theta]}\right) \sim \left(S_\mathcal{T}, Q_{\mathcal{T}}^{[\theta']}\right),\notag\\
	& && \hspace{0.5cm}\forall \theta, \theta' \in [K], \mathcal{T} \subset [N], |\mathcal{T}| \leq T.
\end{align}
That is to say, for any $\theta \in [K]$, the joint distribution of the storage and queries at $T$ or fewer servers are identical.

\noindent\textbf{Answers:} There is a set $\mathcal{E} \subset [N]$ of unresponsive servers and another set $\mathcal{B} \subset [N]$ of Byzantine servers. $\mathcal{B}, \mathcal{E}$ are not necessarily disjoint. The user does not know $\mathcal{E}, \mathcal{B}$ \emph{a priori}, except that
\begin{align}
	|\mathcal{E}| \leq E, |\mathcal{B}| \leq B.
\end{align}
Each server $n \in [N] \setminus (\mathcal{E} \cup \mathcal{B})$  generates the answer $A_n \in [\mathsf{d}]$ using the encoding function $\mathrm{Enc}_{\mathrm{serv}_n}\colon[\mathsf{S}]\times\mathcal{Q} \rightarrow [\mathsf{d}]$ according to its storage and received query, i.e.,
\begin{align}
	A_n = \mathrm{Enc}_{\mathrm{serv}_n}\left(S_n, Q_n^{[\theta]}\right), \forall n \in [N] \setminus (\mathcal{E} \cup \mathcal{B}).
\end{align}
However, any unresponsive or Byzantine server $\bar{n} \in \mathcal{E} \cup \mathcal{B}$ generates an arbitrary answer $A_{\bar{n}} \in [\mathsf{d}]$.

\noindent\textbf{Decoding:} Upon receiving the answers $A_{[N]\setminus\mathcal{E}}$, the user decodes the desired message using a function that depends on $\mathcal{E}$ (since unresponsive servers can be identified by the user), $\mathrm{Dec}_{\mathcal{E}}\colon[K]\times [\mathsf{d}]^{N-|\mathcal{E}|} \times \mathcal{Z}' \rightarrow [\mathsf{M}]$, i.e.,
\begin{align}
	\hat{W} = \mathrm{Dec}_{\mathcal{E}}(\theta, A_{[N]\setminus\mathcal{E}}, Z).
\end{align}

Thus, an E-B-MDS-X-TPIR scheme, is defined as
\begin{align}
	\Psi^C\left(\mathrm{Enc}_{\mathrm{st}},\mathrm{Enc}_{\mathrm{user}},\mathrm{Enc}_{\mathrm{serv}}, \left\{\mathrm{Dec}_{\mathcal{E}}\right\}_{\mathcal{E} \subset [N], |\mathcal{E}| \leq E}\right).
\end{align}

The rate of a classical E-B-MDS-X-TPIR scheme is defined as the number of desired message bits recovered per answer bit that is downloaded from the servers, i.e.,
\begin{align}
	R^C \triangleq \frac{\log(\mathsf{M})}{N\log(\mathsf{d})}.
\end{align}

A rate $R^C$ is said to be achievable if and only if there exists a scheme $\Psi^C$ with this rate such that
\begin{align}
	&\Pr(\hat{W} \neq W_{\theta}) = 0,\notag\\
	&\hspace{1cm}\forall \theta \in [K], \mathcal{E}, \mathcal{B} \subset [N], |\mathcal{E}| \leq E, |\mathcal{B}| \leq B.
\end{align}

\subsection{Quantum Setting}
\noindent\textbf{Shared Entanglement:} In quantum setting, a composite quantum system $\mathcal{A}_{[N]} = \mathcal{A}_1\mathcal{A}_2\cdots\mathcal{A}_{N}$, with underlying Hilbert Space $\mathcal{H}_{\mathcal{A}_{[N]}} = \bigotimes_{n \in [N]}\mathbb{C}^{\mathsf{d}}$ is initialized in the state $\rho_{\mathcal{A}_{[N]}}^{0}$ \emph{a priori}, and the subsystem $\mathcal{A}_n$ is given to server $n, n \in [N]$.

\noindent\textbf{MDS and Secure Storage}: Same as the classical setting.

\noindent\textbf{Queries}: Same as the classical setting.

\noindent\textbf{Answer:} Again, there are unresponsive servers $\mathcal{E}$ and Byzantine servers $\mathcal{B}$ with $\mathcal{E}, \mathcal{B} \subset [N], |\mathcal{E}| \leq E, |\mathcal{B}| \leq B$. Any reliable server applies to its own quantum subsystem a  completely-positive and trace-preserving (CPTP) map as its encoder, based on the realizations of its storage $S_n = s_n$ and received query $Q_n^{[\theta]}=q_n$, i.e.,
\begin{align}
	\mathrm{Enc}_{\mathrm{serv}_n}^{[s_n,q_n]} \colon \mathcal{D}_{\mathcal{A}_n} \rightarrow \mathcal{D}_{\mathcal{A}_n}.
\end{align}
Unresponsive and Byzantine servers apply an arbitrary CPTP map, 
\begin{align}
	\mathcal{M}_{\mathcal{E}\cup\mathcal{B}}\colon\mathcal{D}_{\mathcal{A}_{\mathcal{E}\cup\mathcal{B}}} \rightarrow \mathcal{D}_{\mathcal{A}_{\mathcal{E}\cup\mathcal{B}}}
\end{align}
to their quantum subsystems. Note that Byzantine servers do not change their quantum systems' dimension, as otherwise the user can tell which servers are Byzantine and treat them as erasures instead.

\noindent\textbf{Decoding:}
Upon receiving the quantum system $\mathcal{A}_{[N]\setminus \mathcal{E}}$ with state $\rho_{\mathcal{A}_{[N]\setminus \mathcal{E}}}'$, the user measures with POVM $\mathrm{Dec}_{\mathcal{E}}^{[\theta,z']} = \{\Pi_{\mathcal{E}}^{\theta,z'}(\hat{w}), \hat{w} \in [\mathsf{M}]\}$ that depends on $\theta$ and the realization of its local randomness $Z'=z'$, with outcome random variable $\hat{W}$ as the decoding result.

\begin{remark}
	Both unresponsive and Byzantine servers  apply arbitrary CPTP maps to their quantum systems. The difference is that the indices in $\mathcal{E}$ are directly known to the user after collecting all the answers since unresponsive servers do not respond, while the indices in $\mathcal{B}$ are unknown before decoding. Thus, the decoding POVM $\mathrm{Dec}_{\mathcal{E}}^{[\theta,z']}$ depends on $\mathcal{E}$.
\end{remark}

Thus, an E-B-MDS-X-TPIR scheme, is defined as
\begin{align}
	\Psi^Q\left(\rho^{0},\mathrm{Enc}_{\mathrm{st}},\mathrm{Enc}_{\mathrm{user}},\mathrm{Enc}_{\mathrm{serv}}, \left\{\mathrm{Dec}_{\mathcal{E}}\right\}_{\mathcal{E} \subset [N], |\mathcal{E}| \leq E}\right).
\end{align}

The rate of a Quantum E-B-MDS-X-TPIR scheme is defined as the number of desired message bits recovered per qubit that is downloaded from the servers, i.e.,
\begin{align}
	R^Q \triangleq \frac{\log(\mathsf{M})}{N\log(\mathsf{d})}.
\end{align}

A rate $R^Q$ is said to be achievable if there exists a scheme $\Psi^Q$ with this rate such that for any $\theta \in [K]$,
\begin{align}
	&\Pr(\hat{W} \neq W_{\theta}) = 0,\notag\\
	&\hspace{1cm}\forall \theta \in [K], \mathcal{E}, \mathcal{B} \subset [N], |\mathcal{E}| \leq E, |\mathcal{B}| \leq B.
\end{align}

A (quantum) E-B-MDS-X-TPIR problem is parameterized by $(E,B,K_c,X,T,N,K)$ where $N,K$ are number of servers and messages respectively. We define the following constants for any $E$,$B$,$K_c$,$X$,$T$,$N$,$K$ that will be used throughout this paper, where in the last line of \eqref{eq:constants}, we pick $N+L$ distinct elements in $\Fq$.
\begin{align}
	 & L \triangleq N - (K_c + E + 2B + X + T - 1),\notag                                                                 \\
	 & V \triangleq K_c + X + T - 1, \notag                                                                               \\
	 & \Fq, q = p^r, q \geq L+N,\notag                                                                                    \\
	 & (\bm{\alpha}, \bm{f}) \triangleq (\alpha_1, \cdots, \alpha_N, f_1, \cdots, f_L) \in \Fq^{N+L}.\label{eq:constants}
\end{align}

\begin{remark}
	Since $T$-privacy is for the index $\theta$, and $X$-security is for the shares of messages, both of which are classical even in the quantum setting, quantum analysis is not required while proving the privacy and security of the  quantum protocol. Quantum considerations (e.g., CSS code formalism in Lemma \ref{lem:error_reduction}), are essential only in the proof of \emph{correctness} of the decoding process.
\end{remark}

\section{Main Results}\label{sec:main}
The main result of this paper is a Q-E-B-MDS-X-TPIR scheme/protocol, namely MCSA-CSS. This protocol interprets the classical CSA code based E-B-MDS-X-TPIR scheme of \cite{Jia_Jafar_MDSXSTPIR} in such a way  that desired message symbols appear as ``errors'' added to a code, combines the classical scheme with a CSS code, and decodes the desired message symbols, erasures and Byzantine errors simultaneously through the syndrome decoding of the CSS code. The scheme achieves a higher rate compared with its classical counterpart. As noted, our Q-E-B-MDS-X-TPIR protocol yields the state-of-art achievable rates under the various special cases corresponding to recently studied quantum PIR settings, e.g., Q-B-X-TPIR\cite{Nomeir_Aytekin_Ulukus_QBXSTPIR}, Q-E-TPIR\cite{MH_QSS}, Q-MDS-X-TPIR\cite{Allaix_N_sum_box}, Q-MDS-TPIR\cite{QMDSTPIR}, and Q-TPIR \cite{QTPIR}, without server secrecy constraints. We have the following theorem\footnote{For ease of comparison with quantum PIR problems, similar to \cite{MH_QSS} but unlike \cite{Jia_Jafar_MDSXSTPIR}, the unreceived qudits from unresponsive servers are also counted in the download cost.}, where setting $E=0$ or $B=0$ corresponds to the case of no resilience to erasures or Byzantine servers, respectively, while setting $K_c = 1$ or $X=0$ corresponds to the case of no MDS or $X$-secure storage constraints, respectively.

\begin{theorem}\label{thm:QEB}
	For quantum $K_c$ MDS $X$-secure $T$-private information retrieval with $N$ servers out of which at most $E$ servers are unresponsive and $B$ servers are Byzantine, the rate
	\begin{align}
		& R^Q = \notag                                                    \\
		& \begin{cases}
			  \frac{2(N-E-2B-K_c-X+1)}{N},                                    \\
			  \hspace{0.62cm} (N-E-2B) > (K_c+X+T-1) \geq N/2                 \\
			  \max\left(\frac{N-2E-4B}{N}, \frac{N-E-2B-K_c-X-T+1}{N}\right), \\
			  \hspace{0.62cm} (N-E-2B) \geq N/2 > (K_c+X+T-1)                 \\
			  \frac{N-E-2B-K_c-X-T+1}{N},                                     \\
			  \hspace{0.62cm} N/2 > (N-E-2B) > (K_c+X+T-1)
		  \end{cases}
   \end{align}
	is achievable.
\end{theorem}
\begin{proof}
	The achievability of the third regime $N/2 > (N-E-2B) > (K_c+X+T-1)$ is trivial since a $q$-dimensional qudit can always be used to transmit a classical $q$-ary symbol and the classical scheme in \cite{Jia_Jafar_MDSXSTPIR} can be directly applied. The achievability of the first regime $(N-E-2B) > (K_c+X+T-1) \geq N/2$ will be established by the scheme presented in Section \ref{sec:QEB}.

	The achievability of the second regime, $(N-E-2B) \geq N/2 > (K_c+X+T-1)$, follows from a  combination of the schemes for the first and third regimes, an idea that appears in the preliminary ArXiv version of this paper \cite[Theorem 1, Remark 6]{Lu_Jafar_QEBXSTPIR} and in the subsequent $2^{nd}$ version of \cite{Nomeir_Aytekin_Ulukus_QBXSTPIR}. First of all, $\frac{N-E-2B-K_c-X-T+1}{N}$ is always achievable by the classical scheme. For the achievability of $\frac{N-2E-4B}{N}$, intuitively, when $N/2 > (K_c+X+T-1)$, one can always use the scheme that has more demanding privacy constraints, i.e., the scheme with $\bar{T}$-privacy such that $K_c + X + \bar{T} -1 = N/2$ and $\bar{T} \geq T$. The Q-MDS-X-$\bar{T}$PIR falls into the first regime and the rate can be calculated accordingly. Note that such a choice of $\bar{T}$ needs $N$ to be even so that $N/2$ is an integer. The odd case will be resolved by Remark \ref{rmk:oddN}.
\end{proof}

\begin{remark}
	In the first regime, we note the rate of the quantum scheme is twice of the classical scheme, which matches the maximal superdense coding gain observed thus far in other quantum settings of PIR \cite{QTPIR,QMDSTPIR,Allaix_N_sum_box} (compared with \cite{Sun_Jafar_TPIR,Freij_MDSTPIR} and \cite{Jia_Jafar_MDSXSTPIR} without unresponsive and Byzantine servers), secret sharing \cite{MH_QSS} (compared with \cite{shamir1979share}).
\end{remark}

\section{Classical E-B-MDS-X-TPIR: CSA code}\label{sec:csa}
The classical version of this problem has been studied in \cite{Jia_Jafar_MDSXSTPIR}, and the CSA code based classical scheme there is an essential building block of its quantum version. Let us briefly summarize it here, starting with an example.

\subsection{Example 1: $E=1, B=0, K_c=2, X=1, T=1$ with $N=6$ Servers \cite{Jia_Jafar_MDSXSTPIR}}
Let $L = N - (K_c + E + X + T - 1) = 2$ and $\alpha_1, \cdots, \alpha_{N=6}, f_1, f_{L=2}$ be $8$ distinct elements over $\Fq$ ($q \geq 8$). Let $w_{[K]}$ be the realizations of all the $K$ messages $W_{[K]}$. Each message has $L\times K_c = 4$ symbols from $\Fq$, i.e., for any $k \in [K]$, message $w_k = \{w_k(i,j)\}_{i \in [2], j \in [2]}$ contains $4$ symbols from $\Fq$. Let $\dot{\bw}_{1,1}, \dot{\bw}_{1,2}, \dot{\bw}_{2,1}, \dot{\bw}_{2,2} \in \Fq^{1\times K}$ denote the row vectors that contain the $4$ symbols of the $K$ messages, respectively, i.e.,
\begin{align}
	w_k = \begin{bmatrix}
		      \dot{\bw}_{1,1} \be_{K}^{k} & \dot{\bw}_{1,2} \be_{K}^{k} \\
		      \dot{\bw}_{2,1} \be_{K}^{k} & \dot{\bw}_{2,2} \be_{K}^{k}
	      \end{bmatrix}
\end{align}
where $\be_{K}^{k}$ is the $k^{th}$ column vector of $\bI_K$.

Let the storage randomness $Z = \left\{\bZ_{1,1}, \bZ_{2,1}\right\}$ be uniform over $\Fq^{1\times K} \times \Fq^{1\times K}$ and user randomness $Z' = \left\{\bZ_{l,t}^{\prime(\kappa)}\right\}_{l \in [2], \kappa \in [2], t = 1}$ be uniform over $\left(\Fq^{K\times 1}\right)^{4}$.

\noindent\textbf{Storage:} The storage at server $n$, $n \in [6]$, conditioned on the realization of messages and storage randomness, is $S_n = s_n$ where
\begin{align}
	s_n & =
	\begin{bmatrix}
		s_n(1) &
		s_n(2)
	\end{bmatrix}\notag \\
	    & =
	\bigg[
		\frac{1}{(f_1 - \alpha_n)^2} \dot{\bw}_{1,1} + \frac{1}{f_1 - \alpha_n} \dot{\bw}_{1,2} + \bz_{1,1} \notag\\
		& \hspace{1.5cm}\frac{1}{(f_2 - \alpha_n)^2} \dot{\bw}_{2,1} + \frac{1}{f_2 - \alpha_n} \dot{\bw}_{2,2} + \bz_{2,1}\bigg].
\end{align}
Here $\bz_{1,1}, \bz_{2,1} \in \Fq^{1\times K}$ are the realizations of random vectors $\bZ_{1,1}, \bZ_{2,1}$ respectively. It is not difficult to see that for any $l \in [2]$, $\left(s_1(l), s_2(l), \cdots, s_6(l)\right) l\in[2]$ is a $[6, 3]$ MDS code for $\left(\dot{\bw}_{l,1}, \dot{\bw}_{l,2}, \bz_{l,1}\right)$, and the storage cost at each server is $1/K_c = 1/2$ of the $K$ messages ($s_n(1), s_n(2) \in \Fq^{1\times K}$ while each of $K$ messages contains $4$ symbols from $\Fq$. At the same time we have a secret sharing of $\frac{1}{(\alpha_n - f_1)^2} \dot{\bw}_{l,1} + \frac{1}{\alpha_n - f_1} \dot{\bw}_{l,2}$ with threshold $1$, thus the MDS and $X=1$ security constraint is satisfied.

\noindent\textbf{Queries:} The query generation contains $K_c = 2$ iterations. The query sent from the user to server $n$, $n \in [6]$, conditioned on the realization of the user's local randomness, is $Q_n^{[\theta]} = q_n$ where
\begin{align}
	q_n = \left\{q_n^{(1)}, q_n^{(2)}\right\},
\end{align}
with the superscript indicating the iteration number, and
\begin{align}
	 & q_n^{(1)} = \begin{bmatrix}
		               q_n^{(1)}(1) \\
		               q_n^{(1)}(2)
	               \end{bmatrix}
	=
	\begin{bmatrix}
		(f_1 - \alpha_n)\be_{K}^{\theta} + (f_1 - \alpha_n)^2 \bz_{1,1}^{\prime(1)} \\
		(f_2 - \alpha_n)\be_{K}^{\theta} + (f_2 - \alpha_n)^2 \bz_{2,1}^{\prime(1)}
	\end{bmatrix} \\
	 & q_n^{(2)} = \begin{bmatrix}
		               q_n^{(2)}(1) \\
		               q_n^{(2)}(2)
	               \end{bmatrix}
	=
	\begin{bmatrix}
		\be_{K}^{\theta} + (f_1 - \alpha_n)^2 \bz_{1,1}^{\prime(2)} \\
		\be_{K}^{\theta} + (f_2 - \alpha_n)^2 \bz_{2,1}^{\prime(2)}
	\end{bmatrix}
\end{align}
Here, $\be_{K,\theta}$ is the $\theta^{th}$ column of $\bI_K$, used for choosing the $\theta^{th}$ entry of $\dot{\bw}$, and  $\bz_{l,t}^{\prime(\kappa)} \in \Fq^{K\times 1}, l \in [2], \kappa \in [2], t = 1$ is the realization of corresponding user randomness $\bZ_{l,t}^{\prime(\kappa)}$. It is again not difficult to verify that the queries form secret sharing of $\be_{K}^{\theta}$ with threshold $1$. Thus, the query is $1$-private.

\noindent\textbf{Answer:} The answer generation  takes $K_c = 2$ iterations. Conditioned on the realization of messages, storage and user randomness, the answer sent from server $n$ is $A_n = a_n = \left\{a_n^{(1)}, a_n^{(2)}\right\}$ where in iteration $\kappa \in [2]$, the answer $a_n^{(\kappa)}$ is just a symbol from $\Fq$. Specifically, in the first iteration,
\begin{align}
	a_n^{(1)} & = s_n q_n^{(1)} = s_n(1) q_n^{(1)}(1) + s_n(2) q_n^{(1)}(2)\notag                                                                                                                                       \\
	          & = \frac{1}{f_1 - \alpha_n} \dot{\bw}_{1,1}\be_{K}^{\theta} + \frac{1}{f_2 - \alpha_n} \dot{\bw}_{2,1}\be_{K}^{\theta}\notag                                                                             \\
	          & ~~~+ \left(\dot{\bw}_{1,1}\bz_{1,1}^{\prime(1)} + \dot{\bw}_{2,1}\bz_{2,1}^{\prime(1)} + \dot{\bw}_{1,2}\be_{K}^{\theta} + \dot{\bw}_{2,2}\be_{K}^{\theta}\right)\notag                                 \\
	          & ~~~ +(f_1 - \alpha_n)\left(\dot{\bw}_{1,2}\bz_{1,1}^{\prime(1)} + \bz_{1,1}\be_{K}^{\theta}\right)\notag\\
			  & ~~~ + (f_2 - \alpha_n)\left(\dot{\bw}_{2,2}\bz_{2,1}^{\prime(1)} + \bz_{2,1}\be_{K}^{\theta}\right)\notag \\
	          & ~~~+(f_1 - \alpha_n)^2\bz_{1,1}\bz_{1,1}^{\prime(1)} + (f_2 - \alpha_n)^2\bz_{2,1}\bz_{2,1}^{\prime(1)}                                                                                                 \\
	          & = \frac{1}{f_1 - \alpha_n} \dot{\bw}_{1,1}\be_{K}^{\theta} + \frac{1}{f_2 - \alpha_n} \dot{\bw}_{2,1}\be_{K}^{\theta}\notag\\
			  & ~~~ + * + \alpha_n * + \alpha_n^2 *
\end{align}
where the coefficients for rational terms are desired message symbols and the coefficients for $\alpha_n^0, \alpha_n^1, \alpha_n^2$ are interfering symbols whose specific forms are not important. The collection of the answers from the $6$ servers can be represented as,
\begin{align}
	\begin{bmatrix}
		a_1^{(1)} \\
		a_2^{(1)} \\
		a_3^{(1)} \\
		a_4^{(1)} \\
		a_5^{(1)} \\
		a_6^{(1)}
	\end{bmatrix}
	=
	\begin{bmatrix}
		\frac{1}{f_1 - \alpha_1} & \frac{1}{f_2 - \alpha_1} & 1 & \alpha_1 & \alpha_1^2 \\
		\frac{1}{f_1 - \alpha_2} & \frac{1}{f_2 - \alpha_2} & 1 & \alpha_2 & \alpha_2^2 \\
		\frac{1}{f_1 - \alpha_3} & \frac{1}{f_2 - \alpha_3} & 1 & \alpha_3 & \alpha_3^2 \\
		\frac{1}{f_1 - \alpha_4} & \frac{1}{f_2 - \alpha_4} & 1 & \alpha_4 & \alpha_4^2 \\
		\frac{1}{f_1 - \alpha_5} & \frac{1}{f_2 - \alpha_5} & 1 & \alpha_5 & \alpha_5^2 \\
		\frac{1}{f_1 - \alpha_6} & \frac{1}{f_2 - \alpha_6} & 1 & \alpha_6 & \alpha_6^2 \\
	\end{bmatrix}
	\begin{bmatrix}
		\dot{\bw}_{1,1}\be_{K}^{\theta} \\
		\dot{\bw}_{2,1}\be_{K}^{\theta} \\
		*                               \\
		*                               \\
		*                               \\
	\end{bmatrix}.\label{eq:ans_exp}
\end{align}
Due to the fact that any $5$ rows of the matrix in \eqref{eq:ans_exp} form an invertible sub-matrix according to \cite{Jia_Jafar_MDSXSTPIR}, the answers form a
$[6,5]$ MDS code such that one erasure can be corrected and $2$ desired message symbols $w_{\theta}(:,1) = [\dot{\bw}_{1,1}\be_{K}^{\theta}~~~ \dot{\bw}_{2,1}\be_{K}^{\theta}]^\top$ (together with the interfering symbols) can be decoded.

In the second iteration, the answer from each server is still a symbol in $\Fq$, where
\begin{align}
	a_n^{(2)} & = s_n q_n^{(2)} = s_n(1) q_n^{(2)}(1) + s_n(2) q_n^{(2)}(2)\notag                                                                                                               \\
	          & = \frac{1}{(f_1 - \alpha_n)^2} \dot{\bw}_{1,1}\be_{K}^{\theta} + \frac{1}{(f_2 - \alpha_n)^2} \dot{\bw}_{2,1}\be_{K}^{\theta}\notag                                             \\
	          & ~~~+ \frac{1}{f_1 - \alpha_n} \dot{\bw}_{1,2}\be_{K}^{\theta} + \frac{1}{f_2 - \alpha_n} \dot{\bw}_{2,2}\be_{K}^{\theta}\notag\\
			  & ~~~ + * + \alpha_n * + \alpha_n^2 *.\label{eq:exp_mds_ans}
\end{align}
The details of derivation can be found in \cite{Jia_Jafar_MDSXSTPIR} and are omitted here. Note that the first two terms in \eqref{eq:exp_mds_ans}  are already known from the first iteration of decoding. The $6$ answers together can now be written as,
\begin{align}
	\begin{bmatrix}
		a_1^{(2)} \\
		a_2^{(2)} \\
		a_3^{(2)} \\
		a_4^{(2)} \\
		a_5^{(2)} \\
		a_6^{(2)}
	\end{bmatrix}
	& =
	\begin{bmatrix}
		\frac{1}{f_1 - \alpha_1} & \frac{1}{f_2 - \alpha_1} & 1 & \alpha_1 & \alpha_1^2 \\
		\frac{1}{f_1 - \alpha_2} & \frac{1}{f_2 - \alpha_2} & 1 & \alpha_2 & \alpha_2^2 \\
		\frac{1}{f_1 - \alpha_3} & \frac{1}{f_2 - \alpha_3} & 1 & \alpha_3 & \alpha_3^2 \\
		\frac{1}{f_1 - \alpha_4} & \frac{1}{f_2 - \alpha_4} & 1 & \alpha_4 & \alpha_4^2 \\
		\frac{1}{f_1 - \alpha_5} & \frac{1}{f_2 - \alpha_5} & 1 & \alpha_5 & \alpha_5^2 \\
		\frac{1}{f_1 - \alpha_6} & \frac{1}{f_2 - \alpha_6} & 1 & \alpha_6 & \alpha_6^2 \\
	\end{bmatrix}
	\begin{bmatrix}
		\dot{\bw}_{1,2}\be_{K}^{\theta} \\
		\dot{\bw}_{2,2}\be_{K}^{\theta} \\
		*                               \\
		*                               \\
		*                               \\
	\end{bmatrix}\notag\\
	& ~~~ +
	\underbrace{
		\begin{bmatrix}
			\sum_{l \in [2]}\frac{1}{(f_l - \alpha_1)^2} \dot{\bw}_{l,1}\be_{K}^{\theta} \\
			\sum_{l \in [2]}\frac{1}{(f_l - \alpha_2)^2} \dot{\bw}_{l,1}\be_{K}^{\theta} \\
			\sum_{l \in [2]}\frac{1}{(f_l - \alpha_3)^2} \dot{\bw}_{l,1}\be_{K}^{\theta} \\
			\sum_{l \in [2]}\frac{1}{(f_l - \alpha_4)^2} \dot{\bw}_{l,1}\be_{K}^{\theta} \\
			\sum_{l \in [2]}\frac{1}{(f_l - \alpha_5)^2} \dot{\bw}_{l,1}\be_{K}^{\theta} \\
			\sum_{l \in [2]}\frac{1}{(f_l - \alpha_6)^2} \dot{\bw}_{l,1}\be_{K}^{\theta}
		\end{bmatrix}
	}_{\color{gray}\bm{\sigma}^{(1)}, \mathrm{known}}.
\end{align}
After subtracting $\bm{\sigma}^{(1)}$, $w_{\theta}(:,2) = [\dot{\bw}_{1,2}\be_{K}^{\theta}~~~ \dot{\bw}_{2,2}\be_{K}^{\theta}]^\top$ can be decoded similarly. The $2\times 2$ desired message symbols are retrieved by downloading $6 \times 2$ answer symbols from the servers. The rate achieved is $1/3$.

\subsection{CSA Code for E-B-MDS-X-TPIR}
Recall the constants defined in \eqref{eq:constants}. Each message has $L \times K_c$ symbols from $\Fq$, i.e., for any $k \in [K]$, the realization of message $W_k$, $w_k = \left(w_k(i,j)\right)_{l \in [L], \kappa \in [K_c]}$. Let us define the length-$K$ vector that contains the $(l,\kappa)^{th}$ symbol of all the $K$ messages as
\begin{align}
	\dot{\bw}_{l,\kappa} \triangleq
	\begin{bmatrix}
		w_1(l,\kappa) & w_2(l,\kappa) & \cdots w_K(l,\kappa)
	\end{bmatrix}, \forall l \in [L], \kappa \in [K].\label{eq:message_regroup}
\end{align}
Then for any $k \in [K]$, message $w_k$ can be represented as
\begin{align}
	w_k =
	\begin{bmatrix}
		\dot{\bw}_{1,1}\be_{K}^{k} & \dot{\bw}_{1,2}\be_{K}^{k} & \cdots & \dot{\bw}_{1,K_c}\be_{K}^{k} \\
		\dot{\bw}_{2,1}\be_{K}^{k} & \dot{\bw}_{2,2}\be_{K}^{k} & \cdots & \dot{\bw}_{2,K_c}\be_{K}^{k} \\
		\vdots                     & \vdots                     & \vdots & \vdots                       \\
		\dot{\bw}_{L,1}\be_{K}^{k} & \dot{\bw}_{L,2}\be_{K}^{k} & \cdots & \dot{\bw}_{L,K_c}\be_{K}^{k}
	\end{bmatrix}
	\in \Fq^{L\times K_c}.
\end{align}
The sources of randomness included in this scheme, uniform over their respective alphabet, are as follows,
\begin{align}
	 & Z = \{\bZ_{l,x}\}_{l \in [L], x \in [X]}, \bZ_{l,x} \in \Fq^{K\times 1},\notag                                                                      \\
	 & Z' = \{\bZ_{l,t}^{\prime(\kappa)}\}_{l \in [L], \kappa \in [K_c], t \in [T]}, \bZ_{l,t}^{\prime(\kappa)} \in \Fq^{1\times K}.\label{eq:Zrandomness}
\end{align}
We let $z = \{\bz_{l,x}\}_{l \in [L], x \in [X]}, z' = \{\bz_{l,t}^{\prime(\kappa)}\}_{l \in [L], \kappa \in [K_c], t \in [T]}$ be the realizations.

The CSA scheme in \cite{Jia_Jafar_MDSXSTPIR} is summarized in the following protocol. The specific forms of storage, queries and answers generation functions can be found in Appendix \ref{app:functions}.

\begin{protocol}\label{proto:CSA}
	E-B-MDS-X-TPIR:\\ CSA$\left(\left\{\dot{\bw}_{l,\kappa}\right\}_{l \in [L], \kappa \in [K_c]}, z, z'  \right)$ (Classical)
	\begin{enumerate}
		\item \textbf{Storage:} $s_{[N]} \leftarrow \mathrm{StoreGen}\left(\left\{\dot{\bw}_{l,\kappa}\right\}_{l \in [L], \kappa \in [K_c]}, z\right)$
		\item \textbf{Queries:} $q_{[N]} \leftarrow \mathrm{QueryGen}\left(\theta, z'\right)$
		\item \textbf{Answers:} $a_{[N]} = \left\{a_{[N]}^{(\kappa)}\right\}_{\kappa \in [K_c]} \leftarrow \mathrm{AnsGen}\left(s_{[N]}, q_{[N]}\right)$. Note that for all $\kappa \in [K_c]$, the answers at iteration $\kappa$ are specified in \eqref{eq:CSA_scheme}, 
		\begin{figure*}[htbp]
		      \begin{align}
			      \underbrace{
				      \begin{bmatrix}
					      a_1^{(1)} \\ \vdots \\ a_N^{(1)}
				      \end{bmatrix}}_{\triangleq \ba^{(\kappa)}}
			       & =
			      \underbrace{\left[\begin{array}{ccc|cccc}
						                        \frac{1}{f_1-\alpha_1} & \cdots & \frac{1}{f_L-\alpha_1} & 1      & \alpha_1 & \cdots & \alpha_1^{V-1} \\
						                        \frac{1}{f_1-\alpha_2} & \cdots & \frac{1}{f_L-\alpha_2} & 1      & \alpha_2 & \cdots & \alpha_2^{V-1} \\
						                        \vdots                 & \vdots & \vdots                 & \vdots & \vdots   & \vdots & \vdots         \\
						                        \frac{1}{f_1-\alpha_N} & \cdots & \frac{1}{f_L-\alpha_N} & 1      & \alpha_N & \cdots & \alpha_N^{V-1} \\
					                        \end{array}\right]}_{\triangleq \bG_{\mathrm{CSA}_{N,L,V}^{q,(\bm{\alpha}, \mathbf{f})}}}
			      \underbrace{
			      \begin{bmatrix}
					      \dot{\bw}_{1,\kappa}\be_{K}^{\theta} \\
					      :                                    \\
					      \dot{\bw}_{L,\kappa}\be_{K}^{\theta} \\
					      *                                    \\
					      :                                    \\
					      *
				      \end{bmatrix}}_{=[w_{\theta}(:,\kappa);~ \bm{*}]}
					  +
			      \underbrace{
				      \begin{bmatrix}
					      \sum_{l \in [L], k \in [\kappa-1]}\frac{\dot{\bw}_{l,k}\be_{K}^{\theta}}{(f_l - \alpha_1)^{\kappa - k + 1}} \\
					      \sum_{l \in [L], k \in [\kappa-1]}\frac{\dot{\bw}_{l,k}\be_{K}^{\theta}}{(f_l - \alpha_2)^{\kappa - k + 1}} \\
					      \vdots                                                                                                           \\
					      \sum_{l \in [L], k \in [\kappa-1]}\frac{\dot{\bw}_{l,k}\be_{K}^{\theta}}{(f_l - \alpha_N)^{\kappa - k + 1}} \\
				      \end{bmatrix}}_{\color{gray}\triangleq \bm{\sigma}^{(\kappa-1)},~\mathrm{known}}
			      \label{eq:CSA_scheme}
		      \end{align}
			  \hrule
		\end{figure*}
		\item \textbf{Corrupted Answers:} In each iteration $\kappa \in [K_c]$, the user receives $\hat{\ba}^{(\kappa)}$ (answers from unresponsive servers can be replaced by $0$).
		      \begin{align}
			      \hat{\ba}^{(\kappa)} & = \ba^{(\kappa)} + \bm{\epsilon}_{\mathcal{E}\cup\mathcal{B}}^{(\kappa)} = \notag                                                                                                                         \\
			                           & = \bG_{\mathrm{CSA}_{N,L,V}^{q,(\bm{\alpha}, \mathbf{f})}}[w_{\theta}(:,\kappa);~ \bm{*}] + \bm{\epsilon}_{\mathcal{E}\cup\mathcal{B}}^{(\kappa)} + {\color{gray}\bm{\sigma}^{(\kappa-1)}}  \label{eq:corrupted_CSA}
		      \end{align}
			  where $\mathrm{supp}\left(\bm{\epsilon}_{\mathcal{E}\cup\mathcal{B}}^{(\kappa)}\right) = \mathcal{E}\cup\mathcal{B}$ denotes the errors introduced by unresponsive and Byzantine servers.
		\item \textbf{Decoding:} For each $\kappa \in [K_c]$, the user decodes $w(:,\kappa) = \Phi_{\mathcal{E}}^{\mathrm{CSA}}(\hat{\ba}^{(\kappa)} - {\color{gray}\bm{\sigma}^{(\kappa-1)}})$.
	\end{enumerate}
\end{protocol}

\textbf{In the $1^{st}$ iteration}, $\bm{\sigma}^{(0)} = \bzero$ and the answers from $N$ servers can be regarded as a codeword from $\mathcal{C} = \mathrm{CSA}_{N,L,V}^{q,(\bm{\alpha}, \mathbf{f})}$ code with $\bG_{\mathrm{CSA}_{N,L,V}^{q,(\bm{\alpha}, \mathbf{f})}}$ being the generator matrix, added with errors introduced by unresponsive and Byzantine servers as shown in \eqref{eq:corrupted_CSA}.  The generator matrix is defined in \eqref{eq:CSA_scheme}, and the vector $\bm{*}$ contains $V=K_c+X+T-1$ symbols that are regarded as \emph{interference} that arises due to MDS, security and privacy constraints.  The specific forms of the interference terms are not important. According to the following proposition that states the CSA code is an $[N,L+V]$ MDS code with minimum distance $d = N - (L+V) + 1 \overset{\eqref{eq:constants}}{=} E+2B+1$, the $|\mathcal{E}| \leq E$ erasures and $|\mathcal{B}| \leq B$ Byzantine errors can be corrected and the user is able to recover desired message symbols $w_{\theta}(:,1)$ by the decoding scheme of CSA code $\Phi_{\mathcal{E}}^{\mathrm{CSA}}$.

\begin{proposition}\label{prop:CSA_MDS}
	Any $L + V$ rows of $\bG_{\mathrm{CSA}_{N,L,V}^{q,(\bm{\alpha}, \mathbf{f})}}$ defined in \eqref{eq:CSA_scheme} form an invertible matrix, i.e., $\mathrm{CSA}_{N,L,V}^{q,(\bm{\alpha}, \mathbf{f})} \triangleq \mathrm{colspan}\left(\bG_{\mathrm{CSA}_{N,L,V}^{q,(\bm{\alpha}, \mathbf{f})}}\right)$ code is an $[N, L+V]$ MDS code \cite{{Jia_Jafar_MDSXSTPIR}}.
\end{proposition}

\noindent\textbf{In the $\kappa^{th}$ iteration}, $\kappa \in [K_c]$, the received $N$ answers, after subtracting $\bm{\sigma}^{(\kappa-1)}$ which solely depends on the decoding result in the previous iterations, again form a codeword from $\mathcal{C} = \mathrm{CSA}_{N,L,V}^{q,(\bm{\alpha}, \mathbf{f})}$ code, added with errors. Again, according to Proposition \ref{prop:CSA_MDS}, the user is able to decode $w_{\theta}(:,\kappa)$ in the $\kappa^{th}$ iteration.

The communication rate of the CSA code based scheme is
\begin{align}
	R^C = \frac{K_c L}{K_c N} \overset{\eqref{eq:constants}}{=} \frac{N - E - 2B - K_c - X - T + 1}{N}.
\end{align}

\section{Modified CSA (MCSA) Code}\label{sec:QCSA}
In this section, we propose a Modified CSA (MCSA) Code which is still a classical error correction code, that is intended for classical E-B-MDS-X-TPIR protocol, but more compatible with our eventual Q-E-B-MDS-X-TPIR protocol construction, by turning the RS sub-code of CSA code into a GRS code and leveraging the fact that the dual code of a GRS code is still a GRS code.

\subsection{MCSA Code for E-B-MDS-X-TPIR}
\begin{definition}\label{def:QCSA_G}
	{\bf MCSA Code (Classical)}: A Modified Cross Subspace Alignment code $\mathcal{C} = \mathrm{MCSA}_{N,L,V}^{q,(\bm{\alpha}, \mathbf{f}, \bu)}$ over $\Fq$ is the column space of the generator matrix defined in \eqref{eq:QCSA_G} where $\bG_{\mathrm{CSA}_{N,L,V}^{q,(\bm{\alpha}, \mathbf{f})}}$ is defined in \eqref{eq:CSA_scheme}, $\left(\bm{\alpha}, \mathbf{f}\right) = (\alpha_1, \alpha_2, \cdots, \alpha_N, f_1, f_2, \cdots, f_L)$ are $N+L$ distinct elements and $\bu = (u_1, u_2, \cdots, u_N)$ are $N$ non-zero elements in $\Fq$. By definition, $N \geq L + V$ and $q \geq N+L$.
	\begin{align}
		\bG_{\mathrm{MCSA}_{N,L,V}^{q,(\bm{\alpha}, \mathbf{f}, \bu)}} \triangleq \mathrm{Diag}(\bu) \bG_{\mathrm{CSA}_{N,L,V}^{q,(\bm{\alpha}, \mathbf{f})}}.\label{eq:QCSA_G}
	\end{align}
\end{definition}
The specific form of the generator matrix can be found in \eqref{eq:QCSA_scheme}. For this MCSA code, we have the following proposition.
\begin{proposition}\label{prop:QCSA_MDS}
	Any $L + V$ rows of $\bG_{\mathrm{MCSA}_{N,L,V}^{q,(\bm{\alpha}, \mathbf{f}, \bu)}}$ form an invertible matrix, i.e., $\mathrm{MCSA}_{N,L,V}^{q,(\bm{\alpha}, \mathbf{f},\bu)}$ code is an $[N, L+V]$ MDS code.
\end{proposition}

\begin{proof}
	For any $\mathcal{R} \subset [N], |\mathcal{R}| = L+V$, the $L+V$ rows
	\begin{align}
		\bG_{\mathrm{MCSA}_{N,L,V}^{q,(\bm{\alpha}, \mathbf{f}, \bu)}}(\mathcal{R},:) = \mathrm{Diag}(\bu(\mathcal{R}))\bG_{\mathrm{CSA}_{N,L,V}^{q,(\bm{\alpha}, \mathbf{f})}}(\mathcal{R},:)
	\end{align}
	form an invertible matrix since $\mathrm{Diag}(\bu(\mathcal{R}))$ is invertible as $u_i \neq 0, \forall i \in [N]$, and $\bG_{\mathrm{CSA}_{N,L,V}^{q,(\bm{\alpha}, \mathbf{f})}}$ is invertible according to Proposition \ref{prop:CSA_MDS}.
\end{proof}

Let us specify the form of the \textbf{answers} from an MCSA based classical E-B-MDS-X-TPIR scheme next. Note that for all $\kappa \in [K_c]$, the answers at iteration $\kappa$ are specified in \eqref{eq:QCSA_scheme}.
\begin{figure*}[htbp]
\begin{align}
	\underbrace{
		\begin{bmatrix}
			a_1^{(1)} \\ \vdots \\ a_N^{(1)}
		\end{bmatrix}}_{\triangleq \ba^{(\kappa)}}
	 & =
	\underbrace{
		\left[
			\begin{array}{ccc|cccc}
				\overmat{=\bG_{\mathrm{CRS}^{q,(\bm{\alpha},\mathbf{f},\bu)}_{N,L}}}{\frac{u_1}{f_1 - \alpha_1} & \cdots                                                        & \frac{u_1}{f_L - \alpha_1}}
				                                                                                                & \overmat{=\bG_{\mathrm{GRS}^{q,(\bm{\alpha},\bu)}_{N,V}}}{u_1 & u_1 \alpha_{1}              & \cdots               & u_1 \alpha_{1}^{V-1}}                   \\
				\frac{u_2}{f_1 - \alpha_2}                                                                      & \cdots                                                        & \frac{u_2}{f_L - \alpha_2}
				                                                                                                & u_2                                                           & u_2 \alpha_{2}              & \cdots               & u_2 \alpha_{2}^{V-1}                    \\
				\vdots                                                                                          & \vdots                                                        & \vdots                      & \vdots               & \vdots                & \vdots & \vdots \\
				\frac{u_N}{f_1 - \alpha_N}                                                                      & \cdots                                                        & \frac{u_N}{f_L - \alpha_N}  &
				u_N                                                                                             & u_N \alpha_{N}                                                & \cdots                      & u_N \alpha_{N}^{V-1}
			\end{array}\right]}_{=\bG_{\mathrm{MCSA}_{N,L,V}^{q,(\bm{\alpha}, \mathbf{f}, \bu)}}}
	\underbrace{
	\begin{bmatrix}
			\dot{\bw}_{1,\kappa}\be_{K}^{\theta} \\
			:                                    \\
			\dot{\bw}_{L,\kappa}\be_{K}^{\theta} \\
			*                                    \\
			:                                    \\
			*
		\end{bmatrix}}_{=[w_{\theta}(:,\kappa);~ \bm{*}]}
		+
	\underbrace{
		\begin{bmatrix}
			\sum_{l \in [L], k \in [\kappa-1]}\frac{u_1\dot{\bw}_{l,k}\be_{K}^{\theta}}{(f_l - \alpha_1)^{\kappa - k + 1}} \\
			\sum_{l \in [L], k \in [\kappa-1]}\frac{u_2\dot{\bw}_{l,k}\be_{K}^{\theta}}{(f_l - \alpha_2)^{\kappa - k + 1}} \\
			\vdots                                                                                                              \\
			\sum_{l \in [L], k \in [\kappa-1]}\frac{u_N\dot{\bw}_{l,k}\be_{K}^{\theta}}{(f_l - \alpha_N)^{\kappa - k + 1}} \\
		\end{bmatrix}}_{\color{gray}\triangleq \bm{\sigma}^{(\kappa-1)},~\mathrm{known}}\label{eq:QCSA_scheme}
\end{align}
\hrule
\end{figure*}
\begin{remark}
	For any $\kappa \in [K_c]$, the answers at iteration $\kappa$ specified in \eqref{eq:QCSA_scheme} are equal to the answers in \eqref{eq:CSA_scheme} left-multiplied by the matrix $\mathrm{Diag}(\bu)$. Thus, any CSA based scheme can be easily converted to an MCSA based scheme by letting server $n$ multiply its answer generated from CSA based scheme by $u_n$. The generation of storage and queries remains unchanged. Therefore, $X$-security and $T$-privacy follow from the CSA code based scheme. Meanwhile, the decodability of the desired message is guaranteed by Proposition \ref{prop:QCSA_MDS}, just as the decodability of CSA code based scheme is guaranteed by Proposition \ref{prop:CSA_MDS}.
\end{remark}

\begin{remark}
	Compared with the generator matrix of the code defined in \cite[Eq. (15)]{Allaix_N_sum_box} which is a square matrix, note that the generator matrix in this paper is not square to be able to correct errors introduced by unresponsive and Byzantine servers, and it is an enhanced version of the generator matrix of CSA code in \cite[Eq. (70)]{Jia_Jafar_MDSXSTPIR}.
\end{remark}

The MCSA code based scheme is specified in Protocol \ref{proto:QCSA}. The definition of $\left\{\dot{\bw}_{l,\kappa}\right\}_{l \in [L], \kappa \in [K_c]}$, $z, z'$ are the same as those in \eqref{eq:message_regroup} and \eqref{eq:Zrandomness}. $\bu = (u_1, \cdots, u_N) \in \Fq^N$ are $N$ non-zero elements\footnote{$\bu$ is a constant vector  included in the input to the protocol for ease of executing it twice with different parameters $\bu, \bv$ in the quantum protocol.} in $\Fq$. Again, the storage, queries and answers generation functions are specified in Appendix \ref{app:functions}.

\begin{protocol}\label{proto:QCSA}
	E-B-MDS-X-TPIR: \\ MCSA $\left(\left\{\dot{\bw}_{l,\kappa}\right\}_{l \in [L], \kappa \in [K_c]}, z, z', \bu\right)$ (Classical)

\begin{enumerate}[start=1,wide = 3pt, leftmargin = 0em]	
		\item \textbf{Storage:} $s_{[N]} \leftarrow \mathrm{StoreGen}\left(\left\{\dot{\bw}_{l,\kappa}\right\}_{l \in [L], \kappa \in [K_c]}, z\right)$
		\item \textbf{Queries:} $q_{[N]} \leftarrow \mathrm{QueryGen}\left(\theta, z'\right)$
		\item \textbf{Answers:} $\tilde{a}_{[N]} = \left\{\tilde{a}_{[N]}^{(\kappa)}\right\}_{\kappa \in [K_c]} \leftarrow \mathrm{AnsGen}\left(s_{[N]}, q_{[N]}\right)$, $$a_{[N]}^{(\kappa)} \leftarrow u_n\tilde{a}_{[N]}^{(\kappa)}, \forall n \in [N], \kappa \in [K_c].$$ The $N$ answers  at iteration $\kappa \in [K_c]$ are as follows\footnote{The notation $\underline{w_{\theta}(:,\kappa)} = w_{\theta}(:,\kappa)$  indicates  that this represents a vector.}
		      \begin{align}
			      &\ba^{(\kappa)}  \overset{\eqref{eq:QCSA_scheme}}{=} \bG_{\mathrm{MCSA}^{q,(\bm{\alpha},\mathbf{f},\bu)}_{N,L,V}} [w_{\theta}(:,\kappa);~\bm{*}] + {\color{gray}\bm{\sigma}^{(\kappa-1)}}\notag                                               \\
			                     & \overset{\eqref{eq:QCSA_scheme}}{=}\bG_{\mathrm{GRS}^{q,(\bm{\alpha},\bu)}_{N,V}}\bm{*} + \bG_{\mathrm{CRS}^{q,(\bm{\alpha},\mathbf{f},\bu)}_{N,L}}\underline{w_{\theta}(:,\kappa)} + {\color{gray}\bm{\sigma}^{(\kappa-1)}}\label{eq:answers_QCSA}
		      \end{align}
		\item \textbf{Corrupted Answers:} In each iteration $\kappa \in [K_c]$, the user receives corrupted answers $\hat{\ba}^{(\kappa)}$ (answers from unresponsive servers can be replaced by $0$).
		      \begin{align}
			     & \hat{\ba}^{(\kappa)}  = \ba^{(\kappa)} + \bm{\epsilon}_{\mathcal{E}\cup\mathcal{B}}^{(\kappa)}      \label{eq:corrupted_QCSA}                                                                                                                                                                                                             \\
			                           & = \bG_{\mathrm{GRS}^{q,(\bm{\alpha},\bu)}_{N,V}}\bm{*} + \underbrace{\bG_{\mathrm{CRS}^{q,(\bm{\alpha},\mathbf{f},\bu)}_{N,L}}\underline{w_{\theta}(:,\kappa)} + \bm{\epsilon}_{\mathcal{E}\cup\mathcal{B}}^{(\kappa)}}_{\mbox{``error''}}
										+ {\color{gray}\bm{\sigma}^{(\kappa-1)}}\notag
		      \end{align}
			  where $\mathrm{supp}\left(\bm{\epsilon}_{\mathcal{E}\cup\mathcal{B}}^{(\kappa)}\right) = \mathcal{E}\cup\mathcal{B}$ denotes the errors introduced by unresponsive and Byzantine servers.
		\item \textbf{Decoding:} For any $\kappa \in [K_c]$, user computes the syndrome
		      \begin{align}
			       \bs^{(\kappa)} &\triangleq \bH_{\mathrm{GRS}^{q,(\bm{\alpha},\bu)}_{N,V}}^\top \widehat{\ba}^{(\kappa)}\notag                                                                                                                                                                                                                        \\
			       &= \bH_{\mathrm{GRS}^{q,(\bm{\alpha},\bu)}_{N,V}}^\top \left(\bG_{\mathrm{CRS}^{q,(\bm{\alpha},\mathbf{f},\bu)}_{N,L}}\underline{w_{\theta}(:,\kappa)} + \bm{\epsilon}_{\mathcal{E}\cup\mathcal{B}}^{(\kappa)} \right)\notag\\
				   & ~~~ + {\color{gray}\bH_{\mathrm{GRS}^{q,(\bm{\alpha},\bu)}_{N,V}}^\top\bm{\sigma}^{(\kappa-1)}}.\label{eq:QCSA_syndrome}
		      \end{align}
		      and decodes the desired message through
		      \begin{align}
			    & \Phi_{\mathcal{E}}^{\mathrm{GRS}}\bigg(\bs^{(\kappa)}-\underbrace{{\color{gray}\bH_{\mathrm{GRS}^{q,(\bm{\alpha},\bu)}_{N,V}}^\top\bm{\sigma}^{(\kappa-1)}}}_{\color{gray}\mbox{known}}\bigg) \overset{\eqref{eq:QCSA_syndrome}}{=}\notag\\
				&\Phi_{\mathcal{E}}^{\mathrm{GRS}}\left(\bH_{\mathrm{GRS}^{q,(\bm{\alpha},\bu)}_{N,V}}^\top \left(\bG_{\mathrm{CRS}^{q,(\bm{\alpha},\mathbf{f},\bu)}_{N,L}}\underline{w_{\theta}(:,\kappa)} + \bm{\epsilon}_{\mathcal{E}\cup\mathcal{B}}^{(\kappa)} \right)\right)\notag\\
				&= \left(w_{\theta}(:,\kappa), \bm{\epsilon}_{\mathcal{E}\cup\mathcal{B}}^{(\kappa)}\right),\label{eq:QCSA_syndrome_decoding}
		      \end{align}
			  where $\Phi_{\mathcal{E}}^{\mathrm{GRS}}\colon \Fq^{(N-V)} \rightarrow \Fq^L\times\Fq^N$ is the mapping from the syndrome (after subtracting $\bm{\sigma}^{(\kappa-1)}$ related terms) to the $L$ desired message symbols and the error vector introduced by unresponsive and Byzantine server, when the unresponsive servers are  those with indices in the set $\mathcal{E}$.
	\end{enumerate}
\end{protocol}

\begin{remark}
	Note that besides the difference while generating the answers in step 3, compared with Protocol \ref{proto:CSA}, the interpretation of the answers and user's way of decoding are all different. We will explain these in the following subsection.
\end{remark}

\subsection{MCSA Classical E-B-MDS-X-TPIR--Another Interpretation}
Though the MDS property of the MCSA code guarantees the decodability of message symbols when there are unresponsive and Byzantine servers, in order to make it compatible with the Q-E-B-MDS-X-TPIR scheme based on syndrome measurement of a CSS code, we interpret answers from MCSA code based classical E-B-MDS-X-TPIR scheme as the GRS code of the interfering symbols $*$, with CRS encoded desired message symbols added as ``errors.'' With this interpretation, the decoding of the classical scheme is based on the syndrome decoding of a GRS code.

Specifically, the corrupted answer (after subtracting $\color{gray}\bm{\sigma}$ which is known) in \eqref{eq:corrupted_QCSA} can be interpreted as $\mathrm{GRS}^{q,(\bm{\alpha},\bu)}_{N,V}$ encoded interfering symbols, corrupted by the ``errors" caused by $\mathrm{CRS}$ encoded message symbols, erasures and Byzantine errors. $\bH_{\mathrm{GRS}^{q,(\bm{\alpha},\bu)}_{N,V}} \in \Fq^{N\times (N-V)}$ is the parity check matrix of $\mathrm{GRS}^{q,(\bm{\alpha},\bu)}_{N,V}$, and \eqref{eq:QCSA_syndrome} follows from $\bH^\top\bG = \bzero$.

Next let us prove Lemma \ref{lem:QCSA_syndrome_decoding} which guarantees the existence of the decoding function $\Phi_{\mathcal{E}}^{\mathrm{GRS}}$ in \eqref{eq:QCSA_syndrome_decoding}. Essentially, Lemma \ref{lem:QCSA_syndrome_decoding} says that \emph{all the correctable ``errors'' (including ``errors'' introduced by desired message symbols) have different syndromes}. The ``errors'' introduced by desired messages are similar to erasures in the sense that we know their error basis (columns of $\bG_{\mathrm{CRS}}$). Thus, when $L + E + 2B \overset{\eqref{eq:constants}}{=} N-V = d-1 $ where $L$ is the dimension of the message symbols and $d = N-V+1$ is the minimum distance of the the GRS code, all the ``errors'', including those caused by the desired message symbols, can be decoded from the syndrome.

\begin{lemma}\label{lem:QCSA_syndrome_decoding}
	Let $L + E + 2B = N-V$, as stated in \eqref{eq:constants}. For any given unresponsive servers $\mathcal{E} \subset [N], |\mathcal{E}| \leq E$ and any two sets of Byzantine servers $\mathcal{B}, \mathcal{B}' \subset [N]$, $|\mathcal{B}|, |\mathcal{B}'| \leq B$, the syndromes will differ for any two distinct pairs $(\bw, \bm{\epsilon}_{\mathcal{E}\cup\mathcal{B}}) \neq (\bw', \bm{\epsilon}_{\mathcal{E}\cup\mathcal{B}'}')$, where $\bw, \bw' \in \Fq^{L}$, $\bm{\epsilon}_{\mathcal{E}\cup\mathcal{B}}, \bm{\epsilon}_{\mathcal{E}\cup\mathcal{B}'}' \in \Fq^{N\times 1}$, $\mathrm{supp}(\bm{\epsilon}_{\mathcal{E}\cup\mathcal{B}}) = \mathcal{E}\cup\mathcal{B}$ and $\mathrm{supp}(\bm{\epsilon}_{\mathcal{E}\cup\mathcal{B}'}) = \mathcal{E}\cup\mathcal{B}'$, i.e.,
	\begin{align}
		&\bH_{\mathrm{GRS}^{q,(\bm{\alpha},\bu)}_{N,V}}^\top \left(\bG_{\mathrm{CRS}^{q,(\bm{\alpha},\mathbf{f},\bu)}_{N,L}}\bw + \bm{\epsilon}_{\mathcal{E}\cup\mathcal{B}}\right)\notag\\
		&\neq \bH_{\mathrm{GRS}^{q,(\bm{\alpha},\bu)}_{N,V}}^\top \left(\bG_{\mathrm{CRS}^{q,(\bm{\alpha},\mathbf{f},\bu)}_{N,L}}\bw' + \bm{\epsilon}_{\mathcal{E}\cup\mathcal{B}'}'\right)
	\end{align}
	This implies the existence of the decoding function $\Phi_{\mathcal{E}}^{\mathrm{GRS}} \colon \Fq^{N\times 1} \rightarrow \Fq^{L\times 1} \times \Fq^{N\times 1}$ in \eqref{eq:QCSA_syndrome_decoding}.
\end{lemma}
\begin{proof}
	See Appendix \ref{app:proof_QCSA_syndrome}.
\end{proof}

\subsection{Example 2: $E=0, B=1, K_c=1, X=1, T=1$ with $N=6$ Servers: Protocol \ref{proto:QCSA}}
Let $L = N - (K_c + 2B + X + T - 1) = 2$ and $\alpha_1, \cdots, \alpha_{N=6}, f_1, f_{L=2}$ be $8$ distinct elements over $\Fq$ ($q \geq 8$). Also, let $u_1, u_2, \cdots, u_6$ be $6$ non-zero elements form $\Fq$. Let $w_{[K]}$ be the realizations of all the $K$ messages $W_{[K]}$. Each message has $L\times K_c = 2$ symbols from $\Fq$, i.e., for any $k \in [K]$, message $w_k = \{w_k(i,j)\}_{i \in [2], j = 1}$ contains $2$ symbols from $\Fq$. Let $\dot{\bw}_{1,1}, \dot{\bw}_{2,1}\in \Fq^{1\times K}$ denote the row vectors that contain the $2$ symbols of the $K$ messages, respectively, i.e.,
\begin{align}
	w_k = \begin{bmatrix}
		      \dot{\bw}_{1,1} \be_{K}^{k}\\
		      \dot{\bw}_{2,1} \be_{K}^{k}
	      \end{bmatrix}
\end{align}
where $\be_{K}^{k}$ is the $k^{th}$ column vector of $\bI_K$.

We skip the storage and queries. The (corrupted) answers from the servers have the following representation, where server $2$ is Byzantine so that an error is added to its answer. Note that $V = K_c + X + T - 1 = 2$, and since $K_c = 1$, there is only $\hat{\ba}^{(1)}$ with $\bm{\sigma}^{(0)} = \bzero$.
\begin{align}
	\hat{\ba}^{(1)} &= 
	\begin{bmatrix}
		\hat{a}_1^{(1)} &
		\hat{a}_2^{(1)} &
		\hat{a}_3^{(1)} &
		\hat{a}_4^{(1)} &
		\hat{a}_5^{(1)} &
		\hat{a}_6^{(1)}
	\end{bmatrix}^\top\notag\\
	&=
	\begin{bmatrix}
		\frac{u_1}{f_1 - \alpha_1} & \frac{u_1}{f_2 - \alpha_1} & u_1 & u_1\alpha_1\\
		\frac{u_2}{f_1 - \alpha_2} & \frac{u_2}{f_2 - \alpha_2} & u_2 & u_2\alpha_2\\
		\frac{u_3}{f_1 - \alpha_3} & \frac{u_3}{f_2 - \alpha_3} & u_3 & u_3\alpha_3\\
		\frac{u_4}{f_1 - \alpha_4} & \frac{u_4}{f_2 - \alpha_4} & u_4 & u_4\alpha_4\\
		\frac{u_5}{f_1 - \alpha_5} & \frac{u_5}{f_2 - \alpha_5} & u_5 & u_5\alpha_5\\
		\frac{u_6}{f_1 - \alpha_6} & \frac{u_6}{f_2 - \alpha_6} & u_6 & u_6\alpha_6\\
	\end{bmatrix}
	\begin{bmatrix}
		\dot{\bw}_{1,1}\be_{K}^{\theta} \\
		\dot{\bw}_{2,1}\be_{K}^{\theta} \\
		*                               \\
		*                               
	\end{bmatrix}
	+
	\begin{bmatrix}
		0 \\
		\epsilon \\
		0 \\
		0 \\
		0 \\
		0
	\end{bmatrix}\notag\\
	&=\bG_{\mathrm{GRS}^{q,(\bm{\alpha},\bu)}_{6,2}}\bm{*} + \underbrace{\bG_{\mathrm{CRS}^{q,(\bm{\alpha},\mathbf{f},\bu)}_{6,2}}\underline{w_{\theta}(:,1)} + \bm{\epsilon}_{\{2\}}^{(1)}}_{\mbox{``error''}} 
\end{align}
The error correcting capability of the $[6,2,5]$ GRS code will be utilized to find the two desired message symbols and the server $2$ introduced error $\epsilon$, i.e., the syndrome $\bH_{\mathrm{GRS}^{q,(\bm{\alpha},\bu)}_{6,2}}^\top \hat{\ba}^{(1)}$ uniquely determines the $w_{\theta} = \underline{w_{\theta}(:,1)}$ and $\bm{\epsilon}_{\{2\}}^{(1)}$.

\section{MCSA-CSS Protocol for Q-E-B-MDS-X-TPIR}\label{sec:QEB}
In this section, we propose the MCSA-CSS protocol for the Q-E-B-MDS-X-TPIR problem, based on syndrome measurement of a CSS code, that is constructed from GRS sub-codes of two MCSA codes. Exploiting the fact that the dual code of a GRS code is still a GRS code, a CSS code with $N$ physical qudits can be constructed from two GRS codes. The $N$ physical qudits are then  delivered to $N$ servers.\footnote{Let us clarify that the CSS code is not used to deliver logical qudits to servers. The $N$ physical qudits are initially in a constant pure state and are shared as quantum resources for improving communication efficiency.} Two MCSA codes based classical PIR schemes are executed, and servers  apply Pauli operators to the CSS code according to the answers from the classical scheme. The components of Pauli operators corresponding to the GRS sub-codes of interfering symbols are not detectable, because they commute with the stabilizers. This is due to the fact that the CSS code is constructed from the \emph{same} GRS codes. However, the components associated with the Cauchy RS code encoded message symbols (regarded as ``errors''), along with errors introduced by unresponsive and Byzantine servers, are identified through syndrome measurements.

\subsection{MCSA-CSS Protocol}
The MCSA-CSS scheme is presented as Protocol \ref{proto:QCSA-CSS}. During one execution of the quantum scheme, two independent instances of classical schemes will be executed. Thus, each message has $2LK_c$  symbols from $\Fq$, and the randomness also has twice the size as that in classical cases.

Let $w_{[K]}$ be the realizations of $W_{[K]}$, for any $k \in [K]$. We have $w_k = [w_k^X~~w_k^Z] \in \Fq^{L\times 2K_c}$ where
\begin{align}
	& w_k^X = \left(w_k^X(l,\kappa)\right)_{l \in [L], \kappa \in [K_c]},\notag\\
	& w_k^Z = \left(w_k^Z(l,\kappa)\right)_{l \in [L], \kappa \in [K_c]} \in \Fq^{L\times K_c}
\end{align}
stand for the $X,Z$ parts of message $k$ respectively, so that each part has the same size to a message in the classical case. Similar to \eqref{eq:message_regroup}, define the length-$K$ vector that contains the $(l,\kappa)^{th}$ symbol of all the $K$ messages' $\star$ part ($\star \in \{X,Z\}$)  as
\begin{align}
	\dot{\bw}_{l,\kappa}^{\star} \triangleq [w_1^{\star}(l,\kappa)~~w_2^{\star}(l,\kappa)~~\cdots~~w_K^{\star}(l,\kappa)],\notag\\
	\forall l \in [L], \kappa \in [K_c].
\end{align}
Similarly,  $Z = \{Z^X, Z^Z\}, Z' = \{Z^{\prime X}, Z^{\prime Z}\}$. Each $Z^{\star}, Z^{\prime\star}, \star \in \{X,Z\}$ is specified similarly according to \eqref{eq:Zrandomness} as follows
\begin{align}
	 & Z^{\star} = \{\bZ^{\star}_{l,x}\}_{l \in [L], x \in [X]}, \bZ^{\star}_{l,x} \in \Fq^{K\times 1},\notag                                                                      \\
	 & Z^{\prime\star} = \{\bZ_{l,t}^{\prime(\kappa)\star}\}_{l \in [L], \kappa \in [K_c], t \in [T]}, \bZ_{l,t}^{\prime(\kappa)\star} \in \Fq^{1\times K}.\label{eq:QZrandomness}
\end{align}
Again, let $z,z'$ be their realizations. Let us pick some constants $\bu = (u_1, \cdots, u_N) \in \Fq^N$ s.t. $u_n \neq 0, \forall n \in [N]$. Meanwhile, set $\bv = (v_1, \cdots, v_N)$ as
\begin{align}
	v_i = u_i^{-1}\prod_{j \neq i}(\alpha_i - \alpha_j)^{-1}, \forall i \in [N].\label{eq:dual_GRS}
\end{align}
The protocol is specified as follows. Note that $\bu,\bv$ are constants specified by the protocol.

\begin{protocol}\label{proto:QCSA-CSS}
	Q-E-B-MDS-X-TPIR:\\ MCSA-CSS$\left(\hspace{-2pt}\left\{\dot{\bw}_{l,\kappa}^X, \dot{\bw}_{l,\kappa}^Z\right\}_{l \in [L], \kappa \in [K_c]}, z, z', \bu, \bv\hspace{-2pt}\right)$ (Quantum)
	\begin{enumerate}
		\item \textbf{Share Entanglement:} For all $\kappa \in [K_c]$, $N$ $q$-dimensional qudits $\mathcal{A}_{[N]}^{(\kappa)}$, with initial state
		      \begin{align}
			      &\rho_{\mathcal{A}_{[N]}^{(\kappa)}}^{0} = \ket{\psi}\bra{\psi},\notag\\
				  &\hspace{0.5cm}\ket{\psi} \in \mathrm{CSS}\left(\mathrm{GRS}_{N,V}^{q,(\bm{\alpha}, \bv)}, \mathrm{GRS}_{N,V}^{q,(\bm{\alpha}, \bu)}\right)
		      \end{align}
		      are delivered to $N$ servers so that server $N$ gets $\mathcal{A}_n^{(\kappa)}$. We let $$\rho_{\left\{\mathcal{A}_{[N]}^{(\kappa)}\right\}_{\kappa \in [K_c]}}^0 = \bigotimes_{\kappa \in [K_c]} \rho_{\mathcal{A}_{[N]}^{(\kappa)}}^{0}.$$
		\item \textbf{Storage, Queries, Answers:} Two independent instances (indexed by X and Z) of Protocol \ref{proto:QCSA} will be executed to generate storage, queries and corresponding classical answers. Specifically, execute Protocol \ref{proto:QCSA} with following parameters, so that the storage, queries, and classical answers can be determined by corresponding steps in Protocol \ref{proto:QCSA}, which are, again, generated according to the $3$ functions specified in Appendix \ref{app:functions}.
		\begin{align}
			&\mbox{E-B-MDS-X-TPIR:}\notag\\
			& ~~~ \mbox{MCSA}\left(\left\{\dot{\bw}_{l,\kappa}^{X}\right\}_{l \in [L], \kappa \in [K_c]}, z^X, z^{\prime X}, \bu\right),\notag\\
			&\mbox{E-B-MDS-X-TPIR:}\notag\\
			& ~~~ \mbox{MCSA}\left(\left\{\dot{\bw}_{l,\kappa}^{Z}\right\}_{l \in [L], \kappa \in [K_c]}, z^Z, z^{\prime Z}, \bv\right).
		\end{align}
		For iteration $\kappa \in [K_c]$, the following classical answers are generated according to \eqref{eq:answers_QCSA} where $\ba^{(\kappa)\star} = [a_1^{(\kappa)\star}~~\cdots~~a_N^{(\kappa)\star}]^\top \in \Fq^{N\times 1}$ such that $a_N^{(\kappa)\star}$ is known to server $n$ for $\star \in \{X,Z\}$.
		      \begin{align}
			      \ba^{(\kappa)X} &= \bG_{\mathrm{GRS}^{q,(\bm{\alpha},\bu)}_{N,V}}\bm{*}^X + \bG_{\mathrm{CRS}^{q,(\bm{\alpha},\mathbf{f},\bu)}_{N,L}}\underline{w_{\theta}^{(\kappa)X}(:,\kappa)}\notag\\
				  & ~~~ + {\color{gray}\bm{\sigma}^{(\kappa-1)X}}, \notag \\
			      \ba^{(\kappa)Z} &= \bG_{\mathrm{GRS}^{q,(\bm{\alpha},\bv)}_{N,V}}\bm{*}^Z + \bG_{\mathrm{CRS}^{q,(\bm{\alpha},\mathbf{f},\bv)}_{N,L}}\underline{w_{\theta}^{(\kappa)Z}(:,\kappa)}\notag\\
				  & ~~~ + {\color{gray}\bm{\sigma}^{(\kappa-1)Z}}.\label{eq:two_part_answers}
		      \end{align}
		      Server $n, n \in [N]$ applies $\mathsf{X}^{a_n^{(\kappa)X}}\mathsf{Z}^{a_n^{(\kappa)Z}}$ to qudit $\mathcal{A}_n^{(\kappa)}$ so that the $N$ answer qudits are in the following state \footnote{For ease of analysis, we assume unresponsive or Byzantine servers firstly behave as reliable servers that apply correct Pauli operators to their qudits and then apply a CPTP map $\mathcal{M}$. There is no loss of generality  since any actual CPTP map $\mathcal{M}'$ applied by the unreliable servers can be viewed as a composition of 1) applying correct Pauli operators, 2) reverting the Pauli operators, 3) applying $\mathcal{M}'$ where the composition of the last 2 steps is  $\mathcal{M}$.}.
		      \begin{align}
			      \rho_{\mathcal{A}_{[N]}^{(\kappa)}}^{1} = \left(\mathsf{X}^{\ba^{(\kappa)X}}\mathsf{Z}^{\ba^{(\kappa)X}}\right) \rho_{\mathcal{A}_{[N]}^{(\kappa)}}^{0} \left(\mathsf{X}^{\ba^{(\kappa)X}}\mathsf{Z}^{\ba^{(\kappa)X}}\right)^{\dagger}.
		      \end{align}
		\item \textbf{Corrupted Answers:} For iteration $\kappa \in [K_c]$, the user replaces the unreceived qudits $\mathcal{A}_{\mathcal{E}}^{(\kappa)}$ with $|\mathcal{E}|$ qudits that are in completely mixed state and labels them $\mathcal{A}_{\mathcal{E}}^{(\kappa)}$. The received qudits are in the following state due to the quantum channels applied by unresponsive and Byzantine servers.
		      \begin{align}
			      \rho_{\mathcal{A}_{[N]}^{(\kappa)}}^{2} = \mathrm{id}\otimes\mathcal{M}_{\mathcal{E}\cup\mathcal{B}}(\rho_{\mathcal{A}_{[N]}^{(\kappa)}}^{1})\label{eq:Q_corrupted}
		      \end{align}
		\item \textbf{Decoding:} For each iteration $\kappa \in [K_c]$, the user performs syndrome measurement of\\ $\mathrm{CSS}\left(\mathrm{GRS}_{N,V}^{q,(\bm{\alpha}, \bu)}, \mathrm{GRS}_{N,V}^{q,(\bm{\alpha}, \bv)}\right)$. The  state becomes
		      \begin{align}
			     &  \rho_{\mathcal{A}_{[N]}^{(\kappa)} \mid \bs^{(\kappa)X}, \bs^{(\kappa)Z}}^3 \notag\\
				&   \hspace{0.5cm} = \mathsf{X}^{\hat{\ba}^{(\kappa)X}}\mathsf{Z}^{\hat{\ba}^{(\kappa)Z}}\rho_{\mathcal{A}_n^{(\kappa)}}^0\left(\mathsf{X}^{\hat{\ba}^{(\kappa)X}}\mathsf{Z}^{\hat{\ba}^{(\kappa)Z}}\right)^\dagger,\label{eq:Q_post_syndrome}\\
			&\beforetext{where for $\star \in \{X,Z\}$,}  \hspace{3cm}    \hat{\ba}^{(\kappa)\star} = \ba^{(\kappa)\star} + \bm{\epsilon}_{\mathcal{E}\cup\mathcal{B}}^{(\kappa)\star}\label{eq:two_part_corrupted}
		      \end{align}
		      for some ${\epsilon}_{\mathcal{E}\cup\mathcal{B}}^{(\kappa)\star} \in \Fq^{N\times 1}$.
		      The user obtains the syndrome
		      \begin{align}
			      &\bs^{(\kappa)X} = \bH_{\mathrm{GRS}^{q,(\bm{\alpha},\bu)}_{N,V}}^\top \hat{\ba}^{(\kappa)X}, \notag\\
				  &\bs^{(\kappa)Z} = \bH_{\mathrm{GRS}^{q,(\bm{\alpha},\bv)}_{N,V}}^\top \hat{\ba}^{(\kappa)Z}\label{eq:QCSA_CSS_Syndrome}
		      \end{align}
		      and decodes the desired message symbols through
		      \begin{align}
			       & \Phi_{\mathcal{E}}^{\mathrm{GRS}}\left(\bs^{(\kappa)X}-{\color{gray}\bH_{\mathrm{GRS}^{q,(\bm{\alpha},\bu)}_{N,V}}^\top\bm{\sigma}^{(\kappa-1)X}}\right)\notag\\
					&= \left(w_{\theta}^X(:,\kappa), \bm{\epsilon}_{\mathcal{E}\cup\mathcal{B}}^{(\kappa)X}\right), \notag                 \\
			       & \Phi_{\mathcal{E}}^{\mathrm{GRS}}\left(\bs^{(\kappa)Z}-{\color{gray}\bH_{\mathrm{GRS}^{q,(\bm{\alpha},\bv)}_{N,V}}^\top\bm{\sigma}^{(\kappa-1)Z}}\right)\notag\\
				   & = \left(w_{\theta}^Z(:,\kappa), \bm{\epsilon}_{\mathcal{E}\cup\mathcal{B}}^{(\kappa)Z}\right).\label{eq:rho_decoding}
		      \end{align}
	\end{enumerate}
\end{protocol}

\begin{remark}\label{rmk:oddN}
	We require $V = K_c + X + T -1 \geq N/2$, i.e., interfering symbols occupy at least half of the answer dimensions, so that the CSS code can be constructed from the GRS codes. Consider the second regime of Theorem \ref{thm:QEB}, i.e., $(N-E-2B) \geq N/2 > K_c + X + T -1$, where $N$ is odd. Though it is not possible to find an integer $\bar{T} > T$ such that $K_c + X + \bar{T} - 1 = N/2$, one can find $T_1 > T, T_2 \geq T, T_1 = T_2 + 1$ such that the total interfering dimensions (this idea is also used in the preliminary ArXiv version of this paper \cite[Theorem 1]{Lu_Jafar_QEBXSTPIR} and in the subsequent $2^{nd}$ version of \cite{Nomeir_Aytekin_Ulukus_QBXSTPIR})
	\begin{align}
		K_c+X+T_1-1 + K_c+X+ T_2-1 = N.\label{eq:oddNT}
	\end{align}
	This means that while constructing the two instances of the classical scheme, we have $T_1$ privacy for the X instance, and $T_2$ privacy for the Z instance. By such choice of $T_1, T_2$, during each of $K_c$ iterations, in the first instance, $L_1 = N - E -2B - K_c - X - T_1 + 1$ symbols of desired message are delivered, and in the second instance, $L_2 = N - E -2B - K_c - X - T_2 + 1$ symbols are delivered. Thus, in total $L_1 + L_2 \overset{\eqref{eq:oddNT}}{=} N-2E-4B$ symbols are delivered. The rate $R^Q = (N-2E-4B)/N$ is thus achieved. The key is that the CSS code will be constructed from an $[N,\lfloor N/2\rfloor]$ GRS code and an $[N,\lceil N/2\rceil]$ GRS code.
\end{remark}

Before analyzing the protocol, let us provide an intuitive explanation. The CSS code is constructed based on the GRS sub-codes of two instances of MCSA codes designed for the PIR problem. Since the GRS sub-code corresponds to interfering symbols, the Pauli operators associated with these interfering symbols commute with the stabilizers of the CSS code and, therefore, cannot be detected through syndrome measurement. In contrast, the Pauli operators associated with message symbols, along with any erasures or Byzantine errors, shift the $N$ qudits into an error space that can be uniquely identified through syndrome measurement. In this interpretation, the message symbols act as sources of ``errors." However, since these ``errors" introduced by message symbols have a known basis, they are no more detrimental than erasures. Combined with the fact that a Pauli error corresponds to both $\mathsf{X}$ and $\mathsf{Z}$ errors, each of which can carry classical messages, the CSS code used in Protocol \ref{proto:QCSA-CSS} with minimum distance $d \geq \min(d_X, d_Z) = N-V+1$ can transmit $\bm{2}\bm{L}$ classical symbols, correct $E$ erasures and $B$ Byzantine errors as long as $L + E + 2B \overset{\eqref{eq:constants}}{=} N - V = \min(d_X, d_Z)-1 \leq d-1$.

\subsection{Analysis of MCSA-CSS Protocol}
Let us first prove its correctness.
\subsubsection{Existence of the CSS Code}
According to \cite[(5.1.6) Theorem]{Macwilliams}, with the choice of $\bv$ in \eqref{eq:dual_GRS}, we have
${\mathrm{GRS}_{N,V}^{q,(\bm{\alpha}, \bu)}}^\perp = \mathrm{GRS}_{N,N-V}^{q,(\bm{\alpha}, \bv)} \subset \mathrm{GRS}_{N,V}^{q,(\bm{\alpha}, \bv)}$ when $V \geq N/2$. Thus the $\mathrm{CSS}\left(\mathrm{GRS}_{N,V}^{q,(\bm{\alpha}, \bu)}, \mathrm{GRS}_{N,V}^{q,(\bm{\alpha}, \bv)}\right)$ code exists.

\subsubsection{Corrupted Answers} 
Without loss of generality we assume all the unresponsive and Byzantine servers first apply the correct Pauli Gates as other (reliable) servers, and then apply an arbitrary quantum channel afterwards, as an arbitrary quantum channel can be regarded as a composition of Pauli Gates with another quantum channel.

Recall that we replaced the unreceived qudits with qudits in completely mixed state. This can be viewed as if the unresponsive servers' answer qudits were received but went through a quantum depolarizing channel (Qudit Twirl, \cite[Exercise 4.7.6]{Quantum_Information_Theory}). Thus the state derived in \eqref{eq:Q_corrupted} is correct.

\subsubsection{State after Syndrome Measurement}
The two underlying GRS codes of the CSS code have distance $d_X = d_Z = N-V+1
	\geq N - (L+V) + 1 \overset{\eqref{eq:constants}}{=} E + 2B + 1$, thus $\min(d_X, d_Z) \geq E + 2B + 1$. Thus $|\mathcal{E}\cup\mathcal{B}| \leq E+B \leq \min(d_X, d_Z)-1$, and according to Lemma \ref{lem:error_reduction}, the error reduces to Pauli Operators and \eqref{eq:Q_post_syndrome} is correct.

\subsubsection{Syndrome and Decoding}
Again, according to Lemma \ref{lem:error_reduction}, \eqref{eq:QCSA_CSS_Syndrome} is correct and the decoding reduces to classical case by identifying \eqref{eq:two_part_answers}, \eqref{eq:two_part_corrupted}, \eqref{eq:QCSA_CSS_Syndrome} with \eqref{eq:answers_QCSA}, \eqref{eq:corrupted_QCSA}, \eqref{eq:QCSA_syndrome}. Thus the decodability is guaranteed by the decoder in Protocol \ref{proto:QCSA}.

\subsubsection{MDS, Security and Privacy}
The satisfaction of these constraints is ensured by Protocol \ref{proto:QCSA}, which, in turn, is guaranteed by Protocol \ref{proto:CSA}, as demonstrated in \cite{Jia_Jafar_MDSXSTPIR}. Note that the pre-shared entangled systems do not break the privacy or security since they are completely independent of the messages, randomness, and the index of the desired message.

Finally, consider the rate of the Q-E-B-MDS-X-TPIR scheme in Protocol \ref{proto:QCSA-CSS}. In each iteration $\kappa \in [K_c]$, $N$ qudits are downloaded, and $2L$ desired message symbols $w^{(\kappa)X}(:,\kappa),w^{(\kappa)Z}(:,\kappa)$ are retrieved. Therefore, the overall rate is
\begin{align}
	R^Q = \frac{2K_c L}{K_c N} = \frac{2\left(N - E - 2B - K_c - X - T + 1\right)}{N}.
\end{align}

\section{Conclusion}\label{sec:conclusion}
The Q-E-B-MDS-X-TPIR problem is studied where the main challenge is to find a coding structure that is compatible with $X$-secure, MDS storage, $T$-privacy and the construction of quantum CSS code (MCSA codes), while satisfying erasure and Byzantine error-resilience. The new scheme, MCSA-CSS, leverages the error-correcting capabilities of CSS code to  efficiently encode desired computation results (desired message symbols in the PIR case) into the error space, while correcting quantum erasure and errors. The optimality of the proposed scheme remains a challenging open question. Application of MCSA-CSS to quantum coded distributed computation is a promising direction for future work.

\appendix
\section{Appendix}
\subsection{Proof of Lemma \ref{lem:error_reduction}}\label{app:proof_error_reduction}
While the distance of CSS code can be greater than $\min(d_X, d_Z)$, let us define $\bm{d \triangleq \min(d_X, d_Z)}$ in this proof for ease of notation. Let us prove Lemma \ref{lem:error_reduction} for $\mathcal{S} = [d-1]$. The proof for other realizations of $\mathcal{S}$ follows similarly.

The initial state is $\rho^0 = \mathsf{X}^{\bx}\mathsf{Z}^{\bz}\ket{\psi}\bra{\psi}\left(\mathsf{X}^{\bx}\mathsf{Z}^{\bz}\right)^{\dagger}$. After applying the quantum channel, using the Kraus representation of the channel, we have
\begin{align}
	\rho^1 =\sum_{i} &\left(K_i\otimes\mathsf{X}^{\bzero}\mathsf{Z}^{\bzero}\right)\mathsf{X}^{\bx}\mathsf{Z}^{\bz}\ket{\psi}\notag\\
	&~~~\cdot\bra{\psi}\left(\mathsf{X}^{\bx}\mathsf{Z}^{\bz}\right)^{\dagger}\left(K_i\otimes\mathsf{X}^{\bzero}\mathsf{Z}^{\bzero}\right)^{\dagger}
\end{align}
where $K_i \in \mathbb{C}^{q^{(d-1)} \times q^{(d-1)}}$ and $\bzero$ has length $(n-d+1)$.

Since the $\left\{\mathsf{X}^{\bm{\alpha}}\mathsf{Z}^{\bm{\beta}}\right\}_{\bm{\alpha}, \bm{\beta} \in \Fq^{(d-1) \times 1}}$ form a basis for the linear space of all $q^{(d-1)} \times q^{(d-1)}$ complex matrices \cite{Ketkar06}, by representing $K_i$ as linear combinations of Pauli operators, $\rho^1$ can be further written as
\begin{align}
	\rho^1
	 & = \sum_{\bm{\alpha},\bm{\beta}, \bm{\alpha}',\bm{\beta}' \in \Fq^{(d-1)}}c_{\bm{\alpha},\bm{\beta}}^{\bm{\alpha}',\bm{\beta}'}\left(\mathsf{X}^{\bm{\alpha}}\mathsf{Z}^{\bm{\beta}}\otimes\mathsf{X}^{\bzero}\mathsf{Z}^{\bzero}\right)\mathsf{X}^{\bx}\mathsf{Z}^{\bz}\ket{\psi}\notag\\
	 & \hspace{3.5cm}\cdot\bra{\psi}\left(\mathsf{X}^{\bx}\mathsf{Z}^{\bz}\right)^{\dagger}\left(\mathsf{X}^{\bm{\alpha}'}\mathsf{Z}^{\bm{\beta}'}\otimes\mathsf{X}^{\bzero}\mathsf{Z}^{\bzero}\right)^{\dagger} \notag\\
	 & \overset{\eqref{eq:XZ_mul}}{=} \sum_{\bm{\mu},\bm{\tau}, \bm{\mu}',\bm{\tau}' \in \mathcal{F}}\tilde{c}_{\bm{\mu},\bm{\tau}}^{\bm{\mu}',\bm{\tau}'}~\mathsf{X}^{\bx + \bm{\mu}}\mathsf{Z}^{\bz + \bm{\tau}}\ket{\psi}\bra{\psi}\left(\mathsf{X}^{\bx+\bm{\mu}'}\mathsf{Z}^{\bz+\bm{\tau}'}\right)^{\dagger}\label{eq:CSS_QChannel}
\end{align}
where $c,\tilde{c}$ are some coefficients that depend only on the Kraus Operators, and $\bm{\mu},\bm{\tau}, \bm{\mu}',\bm{\tau}'$ are chosen from
\begin{align}
	\mathcal{F} \triangleq \left\{\bv \in \Fq^{n \times 1} \mid \mathrm{supp}(\bv) = [d-1]\right\}.
\end{align}

After the PVM with orthogonal projections $\left\{\bP_i^{\ba,\bb}\right\}_{i \in \mathbb{F}_p}$ corresponding to  stabilizers $\mathsf{X}^{\ba}\mathsf{Z}^{\bb}$, in \eqref{eq:CSS_QChannel} we have
\begin{align}
	 & \mathsf{X}^{\bx + \bm{\mu}}\mathsf{Z}^{\bz + \bm{\tau}}\ket{\psi}\bra{\psi}\left(\mathsf{X}^{\bx+\bm{\mu}'}\mathsf{Z}^{\bz+\bm{\tau}'}\right)^{\dagger} \longrightarrow \notag                                                                                 \\
	 & \sum_{i \in \mathbb{F}_p}\bP_i^{\ba,\bb} \mathsf{X}^{\bx + \bm{\mu}}\mathsf{Z}^{\bz + \bm{\tau}}\ket{\psi}\bra{\psi}\left(\mathsf{X}^{\bx+\bm{\mu}'}\mathsf{Z}^{\bz+\bm{\tau}'}\right)^{\dagger} {\bP_i^{\ba,\bb}}^{\dagger}.
\end{align}
Note that $\mathsf{X}^{\bx + \bm{\mu}}\mathsf{Z}^{\bz + \bm{\tau}}\ket{\psi}$ and $\mathsf{X}^{\bx + \bm{\mu}'}\mathsf{Z}^{\bz + \bm{\tau}'}\ket{\psi}$ are eigenvectors of all stabilizers, and we have \eqref{eq:measure_projection},
\begin{figure*}
\begin{align}
	 & \bP_i^{\ba,\bb} \mathsf{X}^{\bx + \bm{\mu}}\mathsf{Z}^{\bz + \bm{\tau}}\ket{\psi}\bra{\psi}\left(\mathsf{X}^{\bx+\bm{\mu}'}\mathsf{Z}^{\bz+\bm{\tau}'}\right)^{\dagger} {\bP_i^{\ba,\bb}}^{\dagger} =\notag\\
	 & \begin{cases}
		   \bP_i^{\ba,\bb} \mathsf{X}^{\bx + \bm{\mu}}\mathsf{Z}^{\bz + \bm{\tau}}\ket{\psi}\bzero_{1\times q^n} = \bzero                                          & \mathsf{X}^{\bx + \bm{\mu}}\mathsf{Z}^{\bz + \bm{\tau}}\ket{\psi} \in \mathrm{Im}(\bP_i^{\ba,\bb}), \mathsf{X}^{\bx + \bm{\mu}'}\mathsf{Z}^{\bz + \bm{\tau}'}\ket{\psi} \notin \mathrm{Im}(\bP_i^{\ba,\bb}) \\
		   \bzero_{q^n \times 1}\bra{\psi}\left(\mathsf{X}^{\bx+\bm{\mu}'}\mathsf{Z}^{\bz+\bm{\tau}'}\right)^{\dagger} {\bP_i^{\ba,\bb}}^{\dagger} = \bzero        & \mathsf{X}^{\bx + \bm{\mu}}\mathsf{Z}^{\bz + \bm{\tau}}\ket{\psi} \notin \mathrm{Im}(\bP_i^{\ba,\bb}), \mathsf{X}^{\bx + \bm{\mu}'}\mathsf{Z}^{\bz + \bm{\tau}'}\ket{\psi} \in \mathrm{Im}(\bP_i^{\ba,\bb}) \\
		   \mathsf{X}^{\bx + \bm{\mu}}\mathsf{Z}^{\bz + \bm{\tau}}\ket{\psi}\bra{\psi}\left(\mathsf{X}^{\bx+\bm{\mu}'}\mathsf{Z}^{\bz+\bm{\tau}'}\right)^{\dagger} & \mathsf{X}^{\bx + \bm{\mu}}\mathsf{Z}^{\bz + \bm{\tau}}\ket{\psi}, \mathsf{X}^{\bx + \bm{\mu}'}\mathsf{Z}^{\bz + \bm{\tau}'}\ket{\psi} \in \mathrm{Im}(\bP_i^{\ba,\bb})
	   \end{cases}\label{eq:measure_projection}
\end{align}
\hrule
\end{figure*}
i.e., after measuring with stabilizer $\mathsf{X}^{\ba}\mathsf{Z}^{\bb}$, $\mathsf{X}^{\bx + \bm{\mu}}\mathsf{Z}^{\bz + \bm{\tau}}\ket{\psi}\bra{\psi}\left(\mathsf{X}^{\bx+\bm{\mu}'}\mathsf{Z}^{\bz+\bm{\tau}'}\right)^{\dagger}$ does not disappear if and only if $\mathsf{X}^{\bx + \bm{\mu}}\mathsf{Z}^{\bz + \bm{\tau}}\ket{\psi}$ and $\mathsf{X}^{\bx + \bm{\mu}'}\mathsf{Z}^{\bz + \bm{\tau}'}\ket{\psi}$ lie in the same eigen space of the stabilizer.

Thus, after the syndrome measurement, $\mathsf{X}^{\bx + \bm{\mu}}\mathsf{Z}^{\bz + \bm{\tau}}\ket{\psi}\bra{\psi}\left(\mathsf{X}^{\bx+\bm{\mu}'}\mathsf{Z}^{\bz+\bm{\tau}'}\right)^{\dagger}$ exists if and only if for every stabilizer, $\mathsf{X}^{\bx + \bm{\mu}}\mathsf{Z}^{\bz + \bm{\tau}}\ket{\psi}$ and $\mathsf{X}^{\bx + \bm{\mu}'}\mathsf{Z}^{\bz + \bm{\tau}'}\ket{\psi}$ lie in the same eigen space, or equivalently, they correspond to the same syndrome (similar to Proposition \ref{prop:css_syndrome})
\begin{align}
	& \bH_{\mathcal{C}_Z}^\top \left(\bx + \bm{\mu}\right) = \bH_{\mathcal{C}_Z}^\top \left(\bx + \bm{\mu}'\right)\notag\\
	& \rightarrow \bH_{\mathcal{C}_Z}^\top \left(\bm{\mu} - \bm{\mu}'\right) = \bzero \rightarrow \bm{\mu} = \bm{\mu}', \label{eq:error_reduction_X}     \\
	& \bH_{\mathcal{C}_X}^\top \left(\bz + \bm{\tau}\right) = \bH_{\mathcal{C}_X}^\top \left(\bz + \bm{\tau}'\right)\notag\\
	& \rightarrow \bH_{\mathcal{C}_X}^\top \left(\bm{\tau} - \bm{\tau}'\right) = \bzero \rightarrow \bm{\tau} = \bm{\tau}', \label{eq:error_reduction_Z}
\end{align}
where last the step of \eqref{eq:error_reduction_X} follows from the fact that $\mathrm{wt}(\bm{\mu} - \bm{\mu}') \leq \min(d_X, d_Z) - 1 \leq d_Z - 1$, i.e., $\bm{\mu} - \bm{\mu}' \notin \mathcal{C}_Z = \mathrm{ker}(\bH_{\mathcal{C}_Z})$ if $\bm{\mu} - \bm{\mu}' \neq \bzero$ (the last step of \eqref{eq:error_reduction_Z} follows similarly). After syndrome measurement, the $n$ qudits are in the state,
\begin{align}
	\rho^2 = \sum_{\bm{\mu},\bm{\tau} \in \mathcal{F}}c'_{\bm{\mu},\bm{\tau}}\mathsf{X}^{\bx + \bm{\mu}}\mathsf{Z}^{\bz + \bm{\tau}}\ket{\psi}\bra{\psi}\left(\mathsf{X}^{\bx+\bm{\mu}}\mathsf{Z}^{\bz+\bm{\tau}}\right)^{\dagger}.
\end{align}

Suppose the outcome of syndrome measurement is $\bs_X = \bH_{\mathcal{C}_Z}^\top (\bx+\bm{\epsilon}_{[d-1]}^X), \bs_Z = \bH_{\mathcal{C}_X}^\top (\bz+\bm{\epsilon}_{[d-1]}^Z)$ with $\bm{\epsilon}_{[d-1]}^X, \bm{\epsilon}_{[d-1]}^Z \in \mathcal{F}$. Then $\forall \bm{\mu}, \bm{\tau} \in \mathcal{F}$, the term $\mathsf{X}^{\bx + \bm{\mu}}\mathsf{Z}^{\bz + \bm{\tau}}\ket{\psi}\bra{\psi}\left(\mathsf{X}^{\bx+\bm{\mu}}\mathsf{Z}^{\bz+\bm{\tau}}\right)^{\dagger}$ does not disappear if and only if,
\begin{align}
	\bH_{\mathcal{C}_Z}^\top \left(\bx + \bm{\mu}\right) = \bH_{\mathcal{C}_Z}^\top \left(\bx + \bm{\epsilon}_{[d-1]}^X\right)  & \rightarrow \bm{\mu} = \bm{\epsilon}_{[d-1]}^X, \label{eq:error_fix_X}  \\
	\bH_{\mathcal{C}_X}^\top \left(\bz + \bm{\tau}\right) = \bH_{\mathcal{C}_X}^\top \left(\bz + \bm{\epsilon}_{[d-1]}^Z\right) & \rightarrow \bm{\tau} = \bm{\epsilon}_{[d-1]}^Z, \label{eq:error_fix_Z}
\end{align}
where the correctness of the last steps follows by the same reasoning as that for \eqref{eq:error_reduction_X} and \eqref{eq:error_reduction_Z}. Thus, the state becomes
\begin{align}
	&\rho^3_{ \mid \bs_X,\bs_Z}=\notag\\
	& ~~~ \mathsf{X}^{\bx + \bm{\epsilon}_{[d-1]}^X}\mathsf{Z}^{\bz + \bm{\epsilon}_{[d-1]}^Z}\ket{\psi}\bra{\psi}\left(\mathsf{X}^{\bx+\bm{\epsilon}_{[d-1]}^X}\mathsf{Z}^{\bz+\bm{\epsilon}_{[d-1]}^Z}\right)^{\dagger}.
\end{align}

\subsection{Storage, Queries, Answers Generation in \cite{Jia_Jafar_MDSXSTPIR}}\label{app:functions}
\hspace{-0.45cm}\scalebox{0.84}{
\begin{tcolorbox}[colframe=black, colback=white, coltitle=black, colbacktitle=white, boxrule=0.5pt, sharp corners, title={$\mathrm{StoreGen}\left(\left\{\dot{\bw}_{l,\kappa}\right\}_{l \in [L], \kappa \in [K_c]}, z = \left\{\bz_{l,x}\right\}_{l \in [L], x \in [X]}\right)$}]
	For all $n \in [N], l \in [L]$
	\begin{align}
		 & s_n = [s_n(1) ~~ s_n(2) ~~ \cdots ~~ s_n(L)]\notag                                                                                                                                                \\
		 & s_n(l) = \sum_{l \in [L], \kappa \in [K_c]}\frac{1}{(f_l - \alpha_n)^{K_c - \kappa + 1}}\dot{\bw}_{l,\kappa}\notag\\
		 & \hspace{1cm} + \sum_{x \in [X]} \left(f_l - \alpha_n\right)^{x-1} \bz_{l,x}~~~\in \Fq^{1\times K}
	\end{align}
	Return $s_{[N]}$
\end{tcolorbox}
}\\[0.1cm]
\noindent\scalebox{0.84}{\begin{tcolorbox}[colframe=black, colback=white, coltitle=black, colbacktitle=white, boxrule=0.5pt, sharp corners, title={$\mathrm{QueryGen}\left(\theta, z' = \left\{\bz_{l,t}^{\prime(\kappa)}\right\}_{l \in [L], t \in [T], \kappa \in [K_c]}\right)$}]
	For all $n \in [N], l \in [L]$
	\begin{align}
		 & q_n = \{q_n^{(1)}, q_n^{(2)}, \cdots, q_n^{(K_c)}\}\notag                                                                                                                                                          \\
		 & q_n^{(\kappa)} =
		\begin{bmatrix}
			q_n^{(\kappa)}(1); &
			q_n^{(\kappa)}(2); &
			\cdots;            &
			q_n^{(\kappa)}(L)
		\end{bmatrix}\notag                                                                                                                                                                                                  \\
		 & q_n^{(\kappa)}(l) = \sum_{l \in [L], \kappa \in [K_c]}(f_l - \alpha_n)^{K_c  - \kappa}\be_{K}^{\theta}\\
		 & \hspace{1.2cm} + \sum_{t \in [T]} \left(f_l - \alpha_n\right)^{K_c + t-1} \bz_{l,t}^{\prime(\kappa)}~~~\in \Fq^{K\times 1}\notag
	\end{align}
	Return $q_{[N]}$
\end{tcolorbox}
}\\[0.1cm]
\noindent\scalebox{0.84}{\begin{tcolorbox}[colframe=black, colback=white, coltitle=black, colbacktitle=white, boxrule=0.5pt, sharp corners, title={$\mathrm{AnsGen}\left(s_{[N]}, \left\{q_{[N]}^{(\kappa)}\right\}_{\kappa \in [K_c]}\right)$}]
	For all $n \in [N], \kappa \in [K_c]$
	\begin{align}
		 & a_n = \{a_n^{(1)}, a_n^{(2)}, \cdots, a_n^{(K_c)}\}\notag \\
		 & a_n^{(\kappa)} = s_n q_n^{(\kappa)} ~~~ \in \Fq
	\end{align}
	Return $a_{[N]} = \left\{a_{[N]}^{(\kappa)}\right\}_{\kappa \in [K_c]}$
\end{tcolorbox}
}

\subsection{Proof of Lemma \ref{lem:QCSA_syndrome_decoding}}\label{app:proof_QCSA_syndrome}
We only need to prove 
\begin{align}
	&\bH_{\mathrm{GRS}^{q,(\bm{\alpha},\bu)}_{N,V}}^\top \bigg(\bG_{\mathrm{CRS}^{q,(\bm{\alpha},\mathbf{f},\bu)}_{N,L}}\underbrace{\left(\bw - \bw'\right)}_{\triangleq\bw''} + \underbrace{\left(\bm{\epsilon}_{\mathcal{E}\cup\mathcal{B}} - \bm{\epsilon}_{\mathcal{E}\cup\mathcal{B}'}'\right)}_{\triangleq \bm{\epsilon}_{\mathcal{E}\cup\mathcal{B}\cup\mathcal{B}'}''}\bigg)\notag\\
	&\neq \bzero, ~~~~~~
	\forall (\bw'', \bm{\epsilon}_{\mathcal{E}\cup\mathcal{B}\cup\mathcal{B}'}'') \neq (\bzero, \bzero).
\end{align}

Note that since $|\mathcal{E}\cup\mathcal{B}\cup\mathcal{B}'| \leq E+2B$, we can find a set $\mathcal{S} \subset [N]$ where $|\mathcal{S}| = E+2B$ and $\mathcal{E}\cup\mathcal{B}\cup\mathcal{B}' \subset \mathcal{S}$ so that
\begin{align}
	\bm{\epsilon}_{\mathcal{E}\cup\mathcal{B}\cup\mathcal{B}'}'' = \bI_N(:,\mathcal{S})\bm{\epsilon}'', \bm{\epsilon}'' \in \Fq^{(E+2B) \times 1}.
\end{align}

Thus, we only need to prove for all length-$(L + E + 2B \overset{\eqref{eq:constants}}{=} N-V)$ column vectors $[\bw'';~\bm{\epsilon}''] \neq \bzero$
\begin{align}
	  & \bH_{\mathrm{GRS}^{q,(\bm{\alpha},\bu)}_{N,V}}^\top \left(\bG_{\mathrm{CRS}^{q,(\bm{\alpha},\mathbf{f},\bu)}_{N,L}}\bw'' + \bI_N(:,\mathcal{S})\bm{\epsilon}''\right)
	  \notag         \\
	= & \bH_{\mathrm{GRS}^{q,(\bm{\alpha},\bu)}_{N,V}}^\top \left[\bG_{\mathrm{CRS}^{q,(\bm{\alpha},\mathbf{f},\bu)}_{N,L}}~~~ \bI_N(:,\mathcal{S})\right] \begin{bmatrix}
		                                                                                                                                                       \bw'' \\
		                                                                                                                                                       \bm{\epsilon}''
	                                                                                                                                                       \end{bmatrix} \neq \bzero.
\end{align}

As a consequence, we only need to prove the following $(N-V) \times (N-V)$ matrix is invertible.
\begin{align}
	\bH_{\mathrm{GRS}^{q,(\bm{\alpha},\bu)}_{N,V}}^\top \left[\bG_{\mathrm{CRS}^{q,(\bm{\alpha},\mathbf{f},\bu)}_{N,L}}~~~ \bI_N(:,\mathcal{S})\right]\label{eq:inv_HGI}
\end{align}

For invertibility of \eqref{eq:inv_HGI}, we first prove the following lemma.
\begin{lemma}\label{lem:inv_GGI}
	The following $N\times N$ matrix is invertible.
	\begin{align}
		\left[\bG_{\mathrm{GRS}^{q,(\bm{\alpha},\bu)}_{N,V}}~~~\bG_{\mathrm{CRS}^{q,(\bm{\alpha},\mathbf{f},\bu)}_{N,L}}~~~ \bI_N(:,\mathcal{S})\right]
	\end{align}
\end{lemma}
\begin{proof}
	On  one hand,
	\begin{align}
		 & \forall \bc \neq \bzero \in \mathrm{colspan}\left(\left[\bG_{\mathrm{GRS}^{q,(\bm{\alpha},\bu)}_{N,V}}~~~\bG_{\mathrm{CRS}^{q,(\bm{\alpha},\mathbf{f},\bu)}_{N,L}}\right]\right)\notag\\
		 &\hspace{2cm} \overset{\eqref{eq:QCSA_scheme}}{=} \mathrm{MCSA}^{q,(\bm{\alpha},\mathbf{f},\bu)}_{N,L,V},\notag \\
		 & \mathrm{wt}(\bc) \geq N - (L+V) + 1,
	\end{align}
	since $\mathrm{MCSA}^{q,(\bm{\alpha},\mathbf{f},\bu)}_{N,L,V}$ is an $[N,L+V]$ MDS code according to Proposition \ref{prop:QCSA_MDS}.
	On the other hand,
	\begin{align}
		 & \forall \bc' \neq \bzero \in \mathrm{colspan}(\bI_N(:,\mathcal{S})),\notag \\ 
		 &\mathrm{wt}(\bc') \leq |\mathcal{S}| = \mathrm{rank}\left(\bI_N(:,\mathcal{S})\right) \notag \\
		 & \hspace{1.85cm}= E + 2B \overset{\eqref{eq:constants}}{=}N-(L+V).\\
	\beforetext{Thus,}&	\mathrm{colspan}\left(\left[\bG_{\mathrm{GRS}^{q,(\bm{\alpha},\bu)}_{N,V}}~~~\bG_{\mathrm{CRS}^{q,(\bm{\alpha},\mathbf{f},\bu)}_{N,L}}\right]\right)\notag\\
		& \cap \mathrm{colspan}(\bI_N(:,\mathcal{S})) = \mathrm{colspan}(\bzero).
	\end{align}
	Combined with the following equation
	\begin{align}
		&\mathrm{rank}\left(\left[\bG_{\mathrm{GRS}^{q,(\bm{\alpha},\bu)}_{N,V}}~~~\bG_{\mathrm{CRS}^{q,(\bm{\alpha},\mathbf{f},\bu)}_{N,L}}\right]\right)\notag\\
		& = \mathrm{rank}\left(\bG_{\mathrm{MCSA}^{q,(\bm{\alpha},\mathbf{f},\bu)}_{N,L,V}}\right) = L+V,\notag\\
		&\mathrm{rank}\left(\bI_N(:,\mathcal{S})\right) = E + 2B \overset{\eqref{eq:constants}}{=}N-(L+V)
	\end{align}
	the proof is complete.
\end{proof}

Now let us prove the invertibility of \eqref{eq:inv_HGI} through a contradiction. Suppose to the contrary, the matrix in \eqref{eq:inv_HGI} is not invertible, then there exists $\bv \in \Fq^{(N-V)\times 1}, \bv \neq \bzero$ such that
\begin{align}
	\bH_{\mathrm{GRS}^{q,(\bm{\alpha},\bu)}_{N,V}}^\top\underbrace{\left[\bG_{\mathrm{CRS}^{q,(\bm{\alpha},\mathbf{f},\bu)}_{N,L}}~~~ \bI_N(:,\mathcal{S})\right]\bv}_{\triangleq \bv' \in \Fq^{N\times 1}} = \bzero.
\end{align}
On the one hand, by definition $\bv' \in \mathrm{colspan}([\bG_{\mathrm{CRS}^{q,(\bm{\alpha},\mathbf{f},\bu)}_{N,L}}~~~ \bI_N(:,\mathcal{S})])$. Additionally, $\bv' \neq \bzero$ since $[\bG_{\mathrm{CRS}^{q,(\bm{\alpha},\mathbf{f},\bu)}_{N,L}}~~~ \bI_N(:,\mathcal{S})]$ has rank $N-V$ according to Lemma \ref{lem:inv_GGI} and because $\bv \neq \bzero$. On the other hand, $\bv' \in \mathrm{ker}(\bH_{\mathrm{GRS}^{q,(\bm{\alpha},\bu)}_{N,V}}^\top) = \mathrm{colspan}(\bG_{\mathrm{GRS}^{q,(\bm{\alpha},\bu)}_{N,V}})$.

A contradiction occurs since $\mathrm{colspan}([\bG_{\mathrm{CRS}^{q,(\bm{\alpha},\mathbf{f},\bu)}_{N,L}}~~~ \bI_N(:,\mathcal{S})])$ $\cap \mathrm{colspan}(\bG_{\mathrm{GRS}^{q,(\bm{\alpha},\bu)}_{N,V}}) = \mathrm{colspan}(\bzero) \not\ni\bv'$ according to Lemma \ref{lem:inv_GGI}. Therefore, \eqref{eq:inv_HGI} is invertible.

\bibliographystyle{IEEEtran}
\bibliography{thesis}

\end{document}